\documentclass[aps,prb,twocolumn,floatfix,eqsecnum,showpacs,superscriptaddress]{revtex4-1}
\usepackage{graphicx}
\usepackage{amsmath}
\usepackage{amssymb}
\usepackage{txfonts}
\usepackage{textcomp}
\usepackage{color}
\usepackage{bm}

\begin{document}

\title{Topological Quantum Infidelity}
\author{Chang-Yu Hou}
%\email[]{}
%\thanks{}
\affiliation{Department of Physics, California Institute of Technology, Pasadena, CA 91125}
\affiliation{Department of Physics and Astronomy, University of California at Riverside, Riverside, CA 92521}
%\author{Netanel Lindner}
%\affiliation{Department of Physics, California Institute of Technology, Pasadena, CA 91125}
\author{Gil~Refael}
%\email[]{}
%\thanks{}
\affiliation{Department of Physics, California Institute of Technology, Pasadena, CA 91125}
\affiliation{Institute for Quantum Information and Matter, California Institute of Technology, Pasadena, CA 91125}
\author{Kirill~Shtengel}
%\email[]{kirill.shtengel@ucr.edu}
%\thanks{}
\affiliation{Department of Physics and Astronomy, University of California at Riverside, Riverside, CA 92521}
\affiliation{Institute for Quantum Information and Matter, California Institute of Technology, Pasadena, CA 91125}
\affiliation{Institute for Solid State Physics, University of Tokyo, Kashiwa 277-8581, Japan}
\date{\today}

\begin{abstract}
Can topological quantum entanglement between anyons in one topological medium
``stray'' into a different, topologically distinct medium? In other words,
can quantum information encoded non-locally in the combined state of
non-Abelian anyons be shared between two distinct topological media? We
consider a setup with two $p$-wave superconductors of opposite chirality and
demonstrate that such scenario is indeed possible. The information encoded in
the fermionic parity of two Majorana zero modes, originally within the same
superconducting domain, can be shared between the domains or moved entirely
from one domain to another provided that vortices can tunnel between them in
a controlled fashion.
\end{abstract}

\maketitle

\section{Introduction}
\label{sec:Intro}

The emergence of quasiparticles known as anyons, i.e., particles whose quantum
exchange statistics is neither bosonic nor fermionic
~\cite{Leinaas1977,Wilczek1982a,Wilczek1982b}, is one of the most interesting
collective phenomena in condensed matter systems~\cite{Arovas1984}. An even
more exotic possibility opens whenever a multidimensional degenerate Hilbert
space is associated with several quasiparticles at fixed positions -- these
quasiparticles can potentially obey non-Abelian
statistics~\cite{Moore1991,Nayak1996c,Nayak2008}. In such a case, braiding of
the quasiparticles results in a non-trivial rotation of vector-states in this
multidimensional subspace. As a consequence, the final state of the system
after multiple exchanges depends on their sequence. These properties -- the
multidimensionality of the Hilbert space combined with the braiding operations
which enable transformations of vector-states in this space -- make non-Abelian
anyons a promising platform for quantum
computation~\cite{Kitaev2003,Nayak2008}. The non-local nature of the
computational basis used for encoding quantum information immunizes it from
local perturbations; the discreteness of braiding operations promises
additional robustness of quantum circuitry that relies on them. Before these
conceptual ideas are turned into functioning quantum devices, however, many
aspects of topological quantum architecture must be worked out.

\begin{figure}
\includegraphics[width=8.3cm]{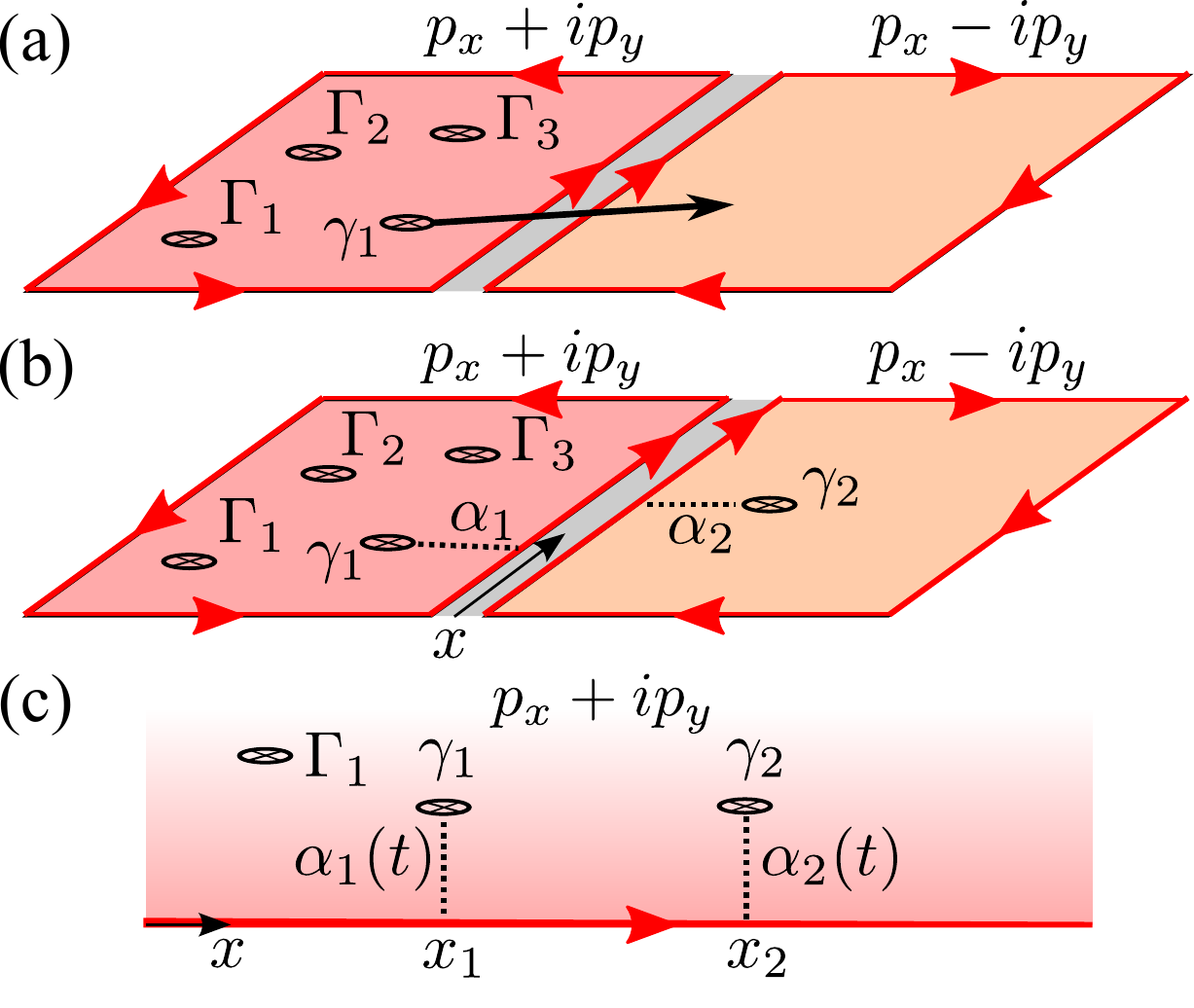}
\caption{(a) An idealized setup involving moving a vortex across the domain wall between
$p_x+ip_y$ and $p_x- i p_y$ superconductors. Red arrows indicate chiral
Majorana edge states at the boundaries of the superconducting domains. $\gamma_1$
denotes a Majorana zero mode associated with the vortex being moved, while $\Gamma_1$,
$\Gamma_2$ and $\Gamma_3$ represent the stationary modes which, along with $\gamma_1$
encode a qubit.
(b) The schematic setup of the model we used to ``mimic'' the move of a vortex
across the domain wall. Majorana modes $\Gamma_1$,\ldots $\Gamma_3$ are
decoupled from the edge states, while $\gamma_1$ and
$\gamma_2$ are coupled to chiral Majorana edge states with
time-dependent coupling constants $\alpha_1(t)$ and $\alpha_2(t)$ at $x=x_1$ and
$x=x_2$ along the edge, respectively. Here, we assume $x_2>x_1$.
(c) A schematic setup considered in Ref.~[\onlinecite{Clarke2011b}]:
two Majorana modes $\gamma_1$ and $\gamma_2$ coupled to a single chiral Majorana
edge with time-dependent coupling constants $\alpha_1(t)$ and $\alpha_2(t)$,
respectively.
}
\label{fig:setup}
\end{figure}

One of the important questions from the computational point of view deals with the
mechanisms for transferring quantum information between different circuit
elements. Ideally, one would imagine that such a transfer happens on-chip
between different qubits (or qudits) defined within the same topological medium
where braiding and measurement operations are employed to manipulate quantum
information. However, this may not always be feasible, particularly if the
complexity of quantum devices were to be scaled up. One possible way of
transferring quantum information would involve more ``conventional'',
non-topological qubits as intermediate
agents~\cite{Hassler2010a,Bonderson2011b,Hassler2011,Pekker2013a,Kovalev2014a}.

An interesting physical question, however, is whether it is possible to
transfer quantum information between distinct topological media directly. An
even more interesting question, at least from the physics point of view, is
whether such information can be \emph{shared} between two such distinct media.
Na\"{i}vely, this seems counterintuitive: topological quantum information is
stored in the (superposition of) degenerate states pertaining to a given
quantum system; sharing such information would require entanglement between two
different quantum systems of macroscopic size. The goal of this paper is to
address these questions. As we show, the answer is, surprisingly, affirmative!

One potentially serious obstacle to this idea is that transferring an anyon
between two distinct media is not only  contingent on an anyon of this type
being supported by both media, it also requires that such a transfer is
accomplished solely by the means of electron tunneling: fractionalized
excitations cannot exist outside of their host media and hence cannot tunnel
between them. E.g., a fractionally charged Laughlin quasiparticle cannot tunnel
between two separate quantum Hall droplets simply due to the fact that each
droplet is made of an integral number of electrons. In  order to avoid these
issues, we focus on a particular example of non-Abelian anyons -- Majorana zero
modes bound to superconducting vortex cores in chiral $p$-wave
superconductors~\cite{Volovik1999,Read2000,Ivanov2001}. For more details on
Majorana zero modes and their potential utility for topological quantum
computation the reader is referred to the recent reviews,
Refs.~[\onlinecite{Alicea2012a}] and [\onlinecite{Beenakker2013a}]. While
recent experimental advances in detecting Majorana zero
modes~\cite{Mourik2012,Deng2012,Das2012a,Rokhinson2012,Nadj-Perge2014} have
followed their theoretical predictions in one-dimensional (1D)
systems~\cite{Kitaev2001,Lutchyn2010a,Oreg2010,Nadj-Perge2013}, in this paper
we focus on two-dimensional (2D) systems instead. 2D chiral superconducting
systems have a distinct conceptual advantage from our point of view: by
considering two superconductors with different chiralities we can be certain
that the two topological regions can never be merged into one, while a
Majorana-hosting vortex can be effectively transferred between the two regions.

An idealized setup considered in this paper is schematically shown in
Fig.~\ref{fig:setup}(a). Quantum information is encoded in the fermionic parity
shared between two Majorana zero modes, $\gamma_1$, $\Gamma_1$. In the standard
four-Majorana qubit encoding~\cite{Bravyi2006}, two additional modes,
$\Gamma_2$ and $\Gamma_3$, serve as a parity reservoir; they do not appear in
the mathematical description of our model in any way. The vortices hosting these four Majorana
modes are initially located within the same superconducting droplet. One of the
vortices is then transferred across the domain wall to a different droplet.
Quantum entanglement is maintained if each of the droplets is individually no
longer in a state of definitive fermionic parity; instead such parity is a
``shared'' property of the droplets.

Intuitively, such a transfer should lead to decoherence of quantum information
as the domain wall separating the two droplets hosts co-propagating gapless
chiral edge states~\cite{Kwon2004,Serban2010}, these states cannot be gapped
and hence even the notion of adiabaticity is not applicable to such a transfer
process. However, as we shall see, these gapless edge states can actually be
used as intermediate agents facilitating quantum information
transfer~\cite{Yao2011}. We should also mention that in this paper we do not
concern ourselves with the actual motion of vortices which host Majorana zero
modes. Not only such a motion can lead to other sources of decoherence (e.g.
via dissipation in the vortex cores), it is not even obvious what a vortex
should look like close to the domain wall between two orthogonal order
parameters; supercurrent-carrying Cooper pairs cannot tunnel between the two
droplets. We circumvent these issues by considering a simplified model where
the locations of all vortices are fixed sufficiently far from the domain wall;
instead the couplings of Majorana zero modes to the edge states are varied as
functions of time -- see Fig.~\ref{fig:setup}(b). When the coupling is weak,
the vortex is effectively well-separated from the edge, and when it is large --
the vortex essentially becomes a part of the
edge~\cite{Rosenow2008a,Rosenow2009a}. By manipulating these couplings, one
vortex can be effectively ``dissolved'' into the edge while another vortex is
``nucleated'' from the edge on the other side.

The paper is organized as follows. In Sec.~\ref{sec:setup}, we present the
minimal model describing the low energy degrees of freedom of our Gedanken
setup consisting of (i) a pair of coupled co-propagating Majorana edge states
at the domain wall and (ii) Majorana zero modes hosted inside the droplets and
coupled to the edge modes with time-dependent coupling strengths. In
Sec.~\ref{sec:QI-transfer}, we show how this setup allows the quantum
information encoded  by Majorana zero modes to be transferred across the domain
wall. We present analytical results obtained within the Heisenberg picture. In
Sec.~\ref{sec:protocol}, the amount of transferred quantum information is
evaluated within a specific protocol governing the time dependence of the
coupling strengths. In Sec.~\ref{sec:discussion}, we discuss the relevant
experimental parameters necessary to achieve high fidelity of such information
transfer. Details of derivations are presented in two Appendices.

\section{Setup}
\label{sec:setup}

\subsection{Domain wall between $p_x \pm i p_y$ superconductors: Co-propagating Majorana edge states}

The presence of a pair of co-propagating Majorana edge states at a domain wall
between $p_x + i p_y$ and $p_x - i p_y$ superconductors, as shown in
Fig.~\ref{fig:setup}, is a direct consequence of the fact that $p_x \pm i p_y$
superconductors are in two distinct topological superconducting phases. As both
$p_x \pm i p_y$ superconductor are fully gapped in the bulk, these chiral
Majorana edge states are the lowest energy degrees of freedom at the domain
wall and are described by the Hamiltonian~\cite{Grosfeld2011}
\begin{multline}
\label{eq:H-two-edges-original}
H_\text{e} = i\int dx \left\{ -  \frac{\hbar v_m}{4} \left[\eta^L (x) \partial_x \eta^L(x)
+  \eta^R (x) \partial_x \eta^R(x) \right]\right.
\\
 + m(x) \eta^{L}(x) \eta^{R}(x) \Bigg\}.
\end{multline}
As shown in Fig.~\ref{fig:setup}(b), $\eta^{L,R}(x)$ are the chiral Majorana
edge fields localized at the left and right sides of the domain wall and $m(x)$
is the tunneling coupling strength between two edge fields. The Majorana edge
fields obey the anti-commutation relation, $\{\eta^{i}(x),\eta^{j}(y)\}= 2
\delta_{ij} \delta(x-y)$, where $i,j=L,R$. Here, we have assumed that both edge
states have the same velocity, $v_m$. We shall discuss the effect of unequal
edge state velocities in Appendix~\ref{app:unequal-velocity}.

Let us consider Majorana modes localized in the vortex cores as depicted in
Fig.~\ref{fig:setup}(b), four of them, $\Gamma_1,\ldots\Gamma_3$ and
$\gamma_1$, are on the left side of the domain wall, while the other one,
$\gamma_2$, is on the right side. All Majorana operators anticommute with one
another and satisfy $\Gamma_a^2=\gamma_a^2=1$. In addition, Majorana operators
anticommute with the operators describing the edge fields. Initially, the
quantum information (fermion parity) is encoded by two Majorana modes,
$\Gamma_1$ and $\gamma_1$, with eigenvalue $i \Gamma_1 \gamma_1(t_0)=\pm 1$ at
initial time $t_0\to -\infty$. Here plus (minus) sign corresponds to odd (even)
parity. (As has been pointed out earlier, two additional modes, $\Gamma_2$ and
$\Gamma_3$, serve as a parity reservoir and do not explicitly enter the
description of our model). In order to transfer the quantum information,
Majorana modes $\gamma_1$ and $\gamma_2$ are coupled to the chiral Majorana
edge field $\eta^{L}(x_1)$ and $\eta^{R}(x_2)$, respectively. On the other
hand, $\Gamma_{a}$'s, the auxiliary Majorana modes needed for encoding the
fermionic parity, are decoupled from the edges and from other zero modes. The
Hamiltonian governing the dynamics of $\gamma_1$ and $\gamma_2$ is then
\begin{multline}
\label{eq:H_c}
H_\text{c} = \frac{i}{2} \int dx \left\{ \alpha_1(t) \eta^L (x) \gamma_1
\delta\left(x-x_1\right)  \right.
\\
\left.
+ \alpha_2(t) \eta^R (x) \gamma_2 \delta\left(x-x_2\right) \right\},
\end{multline}
where the coupling strengths $\alpha_{a}(t)$ for $a \in 1,2$ are
time-dependent; by optimizing this time dependence we can control the transfer
of quantum information.

As a special consequence of the fact that the 1D chiral modes are
co-propagating, the coupling between them (whose strength is given by the
coupling constant $m(x)$ in Eq.~(\ref{eq:H-two-edges-original})) will not
induce a gap. Instead, the coupled edge fields can be diagonalized by a
spatially dependent unitary transformation
\begin{equation}
\label{eq:trans-O}
\left(
\begin{array}{c}
\eta^L(x)\\
\eta^R(x)
\end{array}
\right)
=
\left(
\begin{array}{cc}
\cos \theta(x) & \sin \theta(x)  \\
-\sin \theta(x) & \cos \theta(x)
\end{array}
\right)
\left(
\begin{array}{c}
\eta^1(x)\\
\eta^2(x)
\end{array}
\right),
\end{equation}
where mixing angle $\theta$ depends on the coupling between the two edges and
is given by
\begin{equation}
\label{eq:def-theta_x}
\theta(x) = \theta_0 + \int^{x}_{x_0} dx'\, \frac{2 m(x')}{\hbar v_m}.
\end{equation}
In terms of the ``rotated fields'', the edge Hamiltonian reads
\begin{equation}
\label{eq:H-two-edges-rotated}
H_\text{e} = -\frac{i\hbar v_m}{4}\int dx \,  \left[  \eta^1 (x)  \partial_x \eta^1(x)
+  \eta^2 (x) \partial_x \eta^2(x)\right],
\end{equation}
while the Hamiltonian describing the coupling of the Majorana zero modes to the
edge states becomes
\begin{multline}
H_\text{c}= \frac{i}{2}\int dx \left\{ \left( \lambda^1_1(t) \eta^{1}(x) \gamma_1
+  \lambda^2_1(t) \eta^{2}(x) \gamma_1\right) \delta(x-x_1) \right.
\\
\left.
+  \left( \lambda^1_2(t) \eta^{1}(x) \gamma_2
+  \lambda^2_2(t) \eta^{2}(x) \gamma_2 \right) \delta(x-x_2) \right\}
\label{eq:H_c-new-basis}
\end{multline}
Here, coupling constant $\lambda^{i}_{a} (t)$ with $i,a \in 1,2$ represents the
coupling strength between transformed field $\eta^{i}(x_a)$ and Majorana zero
mode $\gamma_a$. The full set of $\lambda^{i}_{a}$'s is given by
\begin{equation}
\begin{split}
&\lambda^1_1 (t) = + \alpha_1(t) \cos \theta(x_1), \; \lambda^1_2(t) = - \alpha_2(t) \sin \theta(x_2),
\\
&\lambda^2_1(t) =  +\alpha_1(t) \sin \theta(x_1), \; \lambda^2_2(t) = + \alpha_2(t) \cos \theta(x_2) .
\end{split}
\end{equation}

When two coupled chiral edge states have different velocities, the low energy
effective theory can still be described in terms of two independent chiral edge
states~\cite{Wen1991}. The crucial difference now is that these decoupled
Majorana edge states will in general have a non-linear dispersion, in contrast
to the linear dispersion for the case of equal velocities. As a result, there
is no simple spatial-dependent unitary transformation, akin to the one given by
Eq.~\eqref{eq:trans-O}, which decouples the two coupled co-propagating edge
states. However, as long as the velocity difference differnce is small,
${\hbar\delta v}/{m\Delta x}\ll 1$  (where $\Delta x= x_2-x_1$),  the essential
physics is still captured  by the simplified Hamiltonian given by
Eq.~\eqref{eq:H-two-edges-original}, as discussed in
Appendix~\ref{app:unequal-velocity}.

\subsection{Simplified single Majorana edge state setup}
\label{sec:single-edge}

The goal of this paper is to analyze the transfer of quantum entanglement
between two Majorana zero modes, $\gamma_1$ and $\gamma_2$, whose dynamics is
governed by the Hamiltonian $H=H_\text{e}+H_\text{c}$ with $H_\text{e}$ and
$H_\text{c}$ given by Eqs.~\eqref{eq:H-two-edges-rotated} and
\eqref{eq:H_c-new-basis}.  In effect, this Hamiltonian describes two zero modes
simultaneously coupled to two independent chiral edge states. Prior to delving
into this problem, however, let us first study a simplified setup with both
Majorana zero modes coupled to a \emph{single} chiral Majorana edge state as
shown in Fig.~\ref{fig:setup}(c). The Hamiltonian relevant to this setting is
given by
\begin{multline}
\label{eq:H_SE}
H_{\text{SE}} =
\frac{i}{2}\int dx \left\{-\frac{\hbar v_m}{2} \eta(x) \partial_x \eta(x)\right.
\\
+  \alpha_1(t)\eta (x) \gamma_1 \delta(x-x_1)  +  \alpha_2(t) \eta (x) \gamma_2 \delta(x-x_2) \Bigg\},
\end{multline}
where we assume $x_2> x_1$. As shown in Fig.~\ref{fig:setup}(c), we also
consider an auxiliary Majorana zero mode, $\Gamma_1$ which allows us to define
the initial fermion parity, $i \Gamma_1 \gamma_1=\pm 1$. throughout the whole
process, $\Gamma_1$ remains decoupled from the edge state and other zero modes
and hence does not enter the Hamiltonian \eqref{eq:H_SE} which governs the
dynamics of the system. As in the case of two edge states, here the time
dependent coupling strengths $\alpha_a$ control the transfer of quantum
information. We note that employing a chiral edge state of a topological medium
to transport quantum information has been previously discussed in a different
context in Ref.~[\onlinecite{Yao2011}].

In what follows, we first study this simplified setup and establish the
necessary formalism to investigate how the quantum information stored in the
Majorana mode $\gamma_1$ can be transferred to the Majorana mode $\gamma_2$. We
will then apply this formalism to study the original setup described by
Hamiltonian Eqs.~\eqref{eq:H-two-edges-original} and~\eqref{eq:H_c} and discuss
the relation of the original setup and the simplified setup.

\section{Transferring Quantum Information}
\label{sec:QI-transfer}

\subsection{Single Majorana edge state}

A straightforward way to understand the time evolution of quantum states is to employ the Heisenberg picture in quantum mechanics. From Eq.~\eqref{eq:H_SE} together with the commutation relations, the Heisenberg equations of motion of operators are given by
\begin{subequations}
\label{eq:EOM}
\begin{eqnarray}
\label{eq:EOM-eta_SE}
(\partial_t +v_m \partial_x ) \eta &=& \frac{\alpha_1}{\hbar}\gamma_1 \delta(x-x_1) + \frac{\alpha_2}{\hbar} \gamma_2 \delta(x-x_2),
\\
\label{eq:EOM-gamma1_SE}
\partial_t \gamma_1(t)  &=& - \frac{\alpha_1(t)}{\hbar} \eta(x_1,t),
\\
\label{eq:EOM-gamma2_SE}
\partial_t \gamma_2 (t) &=& - \frac{\alpha_2(t)}{\hbar} \eta(x_2,t).
\end{eqnarray}
\end{subequations}
In terms of the initial operators, $ \bar{\gamma}_a\equiv \gamma_a(t_0)$ and
$\eta(x,t_0)$,  the time evolved operators $\gamma_a(t)$ are given by
\begin{subequations}
\label{eq:solution-gamma_a}
\begin{equation}
\label{eq:solution-gamma1_a}
%\begin{split}
\gamma_1(t) = K_{1}(t_0,t)\, \bar{\gamma}_{1}- \int_{t_0}^{t}\! d t'\, \alpha_1(t')\,
K_{1} (t',t)\, \eta^{(0)} (x_1,t'),
\end{equation}
\begin{multline}
\label{eq:solution-gamma2_a}
\gamma_2(t)= K_{2}(t_0,t)\, \bar{\gamma}_{2} +  W(t_0,t,\Delta x)\,\bar{\gamma}_{1}
\\
\qquad \qquad - \int_{t_0}^{t} \! d t'\, \alpha_2(t')\, K_2(t',t)\, \eta^{(1)} (x_2,t').
\end{multline}
%\end{split}
\end{subequations}
Note that, Eq.~\eqref{eq:solution-gamma1_a} contains $\eta^{(0)}(x,t)\equiv
\eta(x-v_m(t-t_0),t_0)$, an unperturbed chiral Majorana edge field. On the
other hand, Eq.~\eqref{eq:solution-gamma2_a} contains $\eta^{(1)}(x,t)$ -- an
edge field already perturbed by the ``upstream'' coupling with $\gamma_1$. The
evolution of the edge field $\eta^{(1)}(x_2,t)$ is given by
\begin{multline}
\label{eq:eta1_x_t}
\eta^{(1)}(x_2,t)= \eta^{(0)}(x,t)- \Theta\left(t-\left(t_0 + \frac{\Delta x}{v_m}
\right) \right) \frac{\alpha_1\!\left(t-{\Delta x}/{v_m}\right)}{\hbar^2 v_m}
\\
\times  \int_{t_0}^{t-\frac{\Delta x}{v_m}} d\tau_0\, \alpha_1(\tau_0)\,
K_1\!\left(\tau_0,t-{\Delta x}/{v_m}\right)\, \eta^{(0)}(x_1,\tau_0).
\end{multline}
A detailed derivation of Eqs.~\eqref{eq:solution-gamma_a} and
\eqref{eq:eta1_x_t} is presented in Appendix~\ref{app:solution-EOM}. The two
functions which determine the time evolution of operators in
Eqs.~\eqref{eq:solution-gamma_a} and \eqref{eq:eta1_x_t} are the kernel
function
\begin{equation}
\label{eq:K_a}
K_{a}(t_\text{i},t_\text{f}) = \exp\left[- \int_{t_\text{i}}^{t_\text{f}} \frac{\alpha_{a}(t')^2}{2 \hbar^2 v_m} dt'\right] ,
\end{equation}
and the weight function
\begin{multline}
\label{eq:W_t}
W(t_0,t,\Delta x) = - \int_{t_0+\frac{\Delta x}{v_m}}^{t} dt'\,
\frac{\alpha_2(t')\, \alpha_1\!\left(t'-{\Delta x}/{v_m}\right)}{\hbar^2 v_m}
\\
\times K_2\!\left(t',t\right)\, K_{1}\!\left(t_0,t'-{\Delta x}/{v_m}\right).
\end{multline}
%where $\Delta x \equiv x_2-x_1$.

Several comments are in order. Firstly, $\gamma_a(t)$ as well as $\eta(x,t)$
satisfy the proper equal-time anticommutation relations, as should be expected
from unitarity. The first terms in Eqs.~\eqref{eq:solution-gamma1_a} and
\eqref{eq:solution-gamma2_a} represent the loss of memory of the initial
conditions for both operators to the coupling of the corresponding zero modes
to the edge.  Most importantly, the second term in the
Eq.~\eqref{eq:solution-gamma2_a} describes how the second Majorana mode
$\gamma_2(t)$ acquires the memory of the initial condition for $\gamma_1(0)$;
the weight function $W(t_0,t,\Delta x)$ given by Eq.~\eqref{eq:W_t} is exactly
the quantum information transferred from the zero mode $\bar{\gamma}_1$ to
$\gamma_2(t)$. Explicitly, let us consider the case where the operator $i
\Gamma_1 \gamma_1(t_0)$ has eigenvalue $\pm 1$ for the initial quantum state.
Using Eq.~\eqref{eq:solution-gamma2_a}, we have $\langle i \Gamma_1 \gamma_2(t)
\rangle=\pm W(t_0,t,\Delta x)$, which represents the fidelity of the quantum
information transfer. In the next section, we will show that owing to the
chiral nature of the edge state, a high fidelity of transfer can be achieved by
employing specific protocols for $\alpha_a(t)$.

As a side note, a non-zero expectation value of $\langle i \gamma_1 \gamma_2(t)
\rangle$ will also develop. Physically, this phenomenon comes from the
edge-state coupling that correlates two Majorana zero modes and can be
understood as contributions from the $\eta^{(0,1)}(x,t)$ terms in
Eqs.~\eqref{eq:solution-gamma_a}. This polarization reaches its peak value of
$2/\pi$ in the limit of $\Delta x\to0$ with the time-independent couplings of
equal strength, $\alpha_1=\alpha_2$~\cite{Rosenow2009a,Clarke2011b}.  Our
approach provides a clear interpretation for the origin of this polarization,
which is discussed in more detail in Appendix~\ref{app:polarization}.

\subsection{Two co-propagating Majorana edge states}
\label{subsec:two-edge}

Having analyzed the case of single edge coupling, we are now ready to discuss
the actual setup of interest shown in Fig.~\ref{fig:setup}(b) where two
Majorana modes are coupled to two co-propagating chiral Majorana edge
states, respectively. The strategy remains the same: starting with the
Hamiltonian given by Eqs.~\eqref{eq:H-two-edges-rotated}
and~\eqref{eq:H_c-new-basis}, to solve the Heisenberg equation of motion, akin
to Eq.~\eqref{eq:EOM}, and to obtain the time evolution of Majorana modes
$\gamma_a(t)$ in the following form
\begin{equation}
\label{eq:solution-gamma_a-2edge}
\begin{split}
\gamma_1(t) =& K_{1}(t_0,t) \bar{\gamma}_{1}+ \dots,
\\
\gamma_2(t)=& K_{2}(t_0,t) \bar{\gamma}_{2} +  W_\text{2E}(t_0,t,\Delta x)\bar{\gamma}_{1} + \dots .
\end{split}
\end{equation}
Here, the $\dots$ represents  contributions from $\eta^{L,R}(x,t_0)$. As those
terms play no part in the quantities we are interested in, we will omit them
for simplicity. Again, $\bar{\gamma}_a\equiv \gamma_a(t_0)$ represent initial
operators while the kernel function $K_a$ is defined in Eq.~\eqref{eq:K_a}.

The weight function, which describes the transfer of the quantum entanglement
across the domain wall, can be related to that of the single Majorana edge
state in Eq.~\eqref{eq:W_t} by
\begin{equation}
W_\text{2E}(t_0,t,\Delta x) = W(t_0,t,\Delta x) \sin \Delta \theta.\label{weightW}
\end{equation}
The phase factor $\Delta \theta$ depends on the coupling strength $m(x)$
between two chiral Majorana edge states and is given by
\begin{equation}
\label{eq:Delta-theta}
\Delta \theta = - \int^{x_2}_{x_1} dx' \frac{2 m(x')}{\hbar v_m}.
\end{equation}
The reduction of the fidelity for quantum information transfer due to this
phase factor can be intuitively understood as follows. The coupling $m(x)$
between two Majorana edge states spatially interchanges contents of the edge
fields $\eta^{L,R}(x)$ in terms of the mixed edge fields $\eta^{1,2}(x)$. As a
result, the quantum entanglement associated with  the initial operator
$\bar{\gamma}_1$ carried away by the edge field $\eta^{L}(x)$ from $x=x_1$ is
partially transferred across the domain wall with the ratio $\sin \Delta
\theta$ and is then picked up by the $\gamma_2$ coupled to the edge field
$\eta^{R}(x)$ at $x=x_2$.

\section{Time-dependent protocol and quantum information transfer}
\label{sec:protocol}

To mimic the process of moving a vortex across the domain wall, the
time-dependent profiles for coupling strengths $\alpha_a(t)$ are chosen to
satisfy the following boundary conditions. At the initial time $t\to -\infty$,
Majorana mode $\gamma_1$ is decoupled from the edge while Majorana mode
$\gamma_2$ is strongly coupled to the edge. At the end of the process, $t\to
\infty$, the Majorana mode $\gamma_1$ is strongly coupled to the edge while the
mode $\gamma_2$ becomes decoupled from the edge. We choose to parameterize the
time dependence of the coupling strengths as
\begin{equation}
\label{eq:alpha-profiles}
\alpha_1 (t) =\frac{\Lambda}{2} ( 1+\tanh \beta t ), \quad \alpha_2 (t)
=\frac{\Lambda}{2} ( 1- \tanh \beta (t-\Delta t) ),
\end{equation}
which satisfy the aforementioned boundary conditions. Here, $\Lambda$
determines the strength of the coupling between the Majorana modes in the bulk
and the edge states, the inverse time constant $\beta$ controls how fast the
couplings are turned on and off, and $\Delta t$ is the time delay between the
two processes.

\begin{figure}
\includegraphics[width=8.3cm]{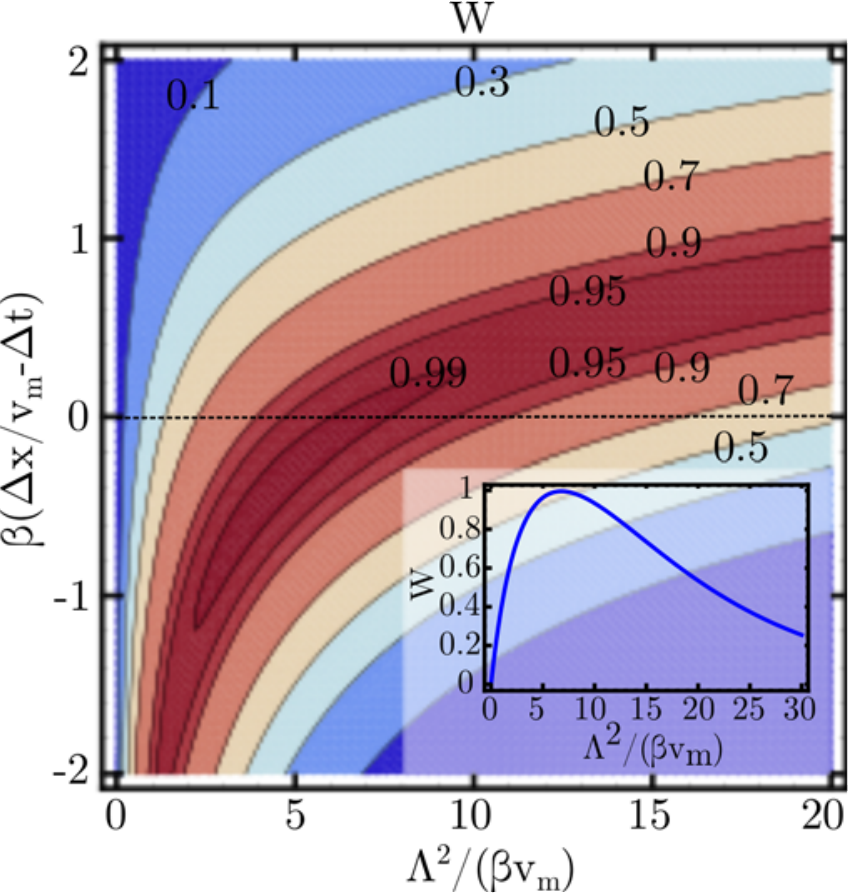}
\caption{The contour plot shows the weight $W(t_0,t,\Delta x)$
as a function of dimensionless parameters, $\Lambda^2/\beta \hbar^2 v_m$ and
$\beta(\Delta x/v_m -\Delta t)$. The weight function is given in
Eq.~\eqref{eq:W_t} and is evaluated using the time-dependent coupling strengths given
in Eq.~\eqref{eq:alpha-profiles}. The inset shows the  weight function $W$
along the line cut $\beta(\Delta x/v_m -\Delta t)=0$ as indicated by the dashed
horizontal line.
}
\label{fig:weight}
\end{figure}

With these time-dependent coupling constants, the weight function
$W(t_0,t,\Delta x)$ in Eq.~\eqref{eq:W_t} is governed by two dimensionless
parameters, $\Lambda^2/\beta \hbar^2 v_m$ and $\beta(\Delta x/v_m -\Delta t)$.
We evaluate the weight function with time $t\to \infty$ and plot the weight
strength as functions of these two dimensionless parameters in
Fig.~\ref{fig:weight}. When both Majorana modes couple to single Majorana edge,
the quantum information transferred from Majorana mode $\gamma_1$ to $\gamma_2$
is fully determined by this weight strength. From Fig.~\ref{fig:weight}, one
can easily achieve more than $95\%$ of fidelity for quantum information
transferring. The inset of Fig.~\ref{fig:weight} shows the weight function
$W(t_0,t \to \infty,\Delta x)$ along the line $\beta(\Delta x/v_m -\Delta
t)=0$. The peak value of this function is about $99.6\%$ and can be achieved by
tuning the ratio of $\Lambda^2/(v_m \hbar^2 \beta)$. I.e., there is a range of
parameters allowing for a reasonable amount of quantum information to be
transferred across the domain wall.

The fidelity of the information transfer across a chiral $p$-wave domain wall,
can be further reduced by  the factor $\sin \Delta \theta$ -- see
Eq.~\eqref{weightW}). However, as we discuss in the next section, this problem
can be in principle alleviated, resulting in an optimal quantum information
transfer.

\section{Discussion}
\label{sec:discussion}

Let us now discuss the experimental parameters pertinent to our setup and their
tunability for optimizing the transfer of quantum information across the domain
wall separating two $p$-wave superconductors with opposite chiralities. To
reach the optimal fidelity, the accumulated phase in Eq.~\eqref{eq:Delta-theta}
due to the coupling between chiral Majorana edge states has to be $\Delta
\theta= \pi/2$ (mod $\pi$), while the weight function $W(t_0,t\to \infty,\Delta
x)$ should be at its maximum.

In order to satisfy the former condition, we observe that the coupling strength
$m(x)$ relates to the superconducting phase difference $\Delta \phi(x)$ across
the domain wall by $m(x)=m_0(x) \cos(\Delta
\phi(x)/2)$.~\cite{Kwon2004,Serban2010} Here $m_0(x)$ is the bare coupling
strength along the domain in the absence of the phase difference. As threading
a magnetic field through the domain wall effectively changes the phase
difference $\Delta \phi(x)/2$, it effectively tune the coupling strength
$m(x)$. Hence, the accumulated phase can be adjusted to $\Delta \theta= \pi/2$
(mod $\pi$). With the estimated value of the bare coupling strength $m_0\sim
0.025\,$meV and the typical edge velocity $v_m\sim
10^5\,$m/s,~\cite{Grosfeld2011} the accumulated phase is estimated to be
$\Delta \theta \sim \Delta x \times 10^6 \,${\textmu}m$^{-1}$. This implies
that $\Delta x$ should be of order $1 - 2\,${\textmu}m in order to have $\Delta
\theta= \pi/2$.

On the other hand, the weight strength $W(t_0,t\to \infty,\Delta)$ can be tuned
by $\beta$, $\Lambda$, and $\Delta t$. Assuming that the inverse time constant
is in the range of $\beta\sim 10^6$ -- $10^9$ s$^{-1}$, it should be easy to
achieve sufficiently small values of $\beta(\Delta x/v_m-\Delta t)$ when
$\Delta x$ is in the order of $1-2\,${\textmu}m and edge velocity $v_m\sim
10^5$m/s. With this condition, the weight function reaches its peak value with
the ratio $\Lambda^2/(v_m \hbar^2 \beta)\sim 7$, as seen in the inset of
Fig.~\ref{fig:weight}. This requires the value of $\Lambda \sim 5 \times
10^{-10}$ -- $2\times 10^{-8}$ eV$\cdot$m$^{1/2}$. Again, we should emphasis
that the actual values of $\beta$ and $\Lambda$ are not essential. Instead, it
is their combination, $\Lambda^2/(v_m \hbar^2 \beta)$, that is crucial for
determining the amount of quantum information transferred.

It is also worth mentioning that while we have studied a specific protocol
given by Eq.~\eqref{eq:alpha-profiles}, other protocols may lead to even more
reliable transfer of quantum entanglement.

\section{Conclusion}
\label{sec:conclusion}

We have described a process whereby quantum entanglement associated with
Majorana zero modes in a chiral $p$-wave superconductor can stray from the
original medium into another, topologically distinct medium. As a result, the
topological quantum information initially encoded within one superconducting
domain can now be shared between two domains of opposite chirality. In
principle, this can be done very efficiently provided that the fermion
tunneling between superconducting vortices and edge states can be judiciously
controlled. From the physical point of view, this effect can be understood from
the fact that the quantum information is encoded in the fermionic parity.
Different domains, even though topologically distinct, need not be in the
parity eigenstates individually; coupling their respective edge states through
fermionic tunneling  provides a mechanism for moving this parity across and
creating entangled states between the domains. Utilizing this mechanism
repeatedly, quantum information can be transferred from one medium to another
directly, without any need for measuring it or employing intermediate quantum
buses.

The reliance of our proposed setup on fermionic tunneling does appear to be a
limiting factor as far as potential generalizations of this scheme to other
topological platforms are concerned. Unfortunately, it is not clear to us how
this constraint can be physically avoided in the cases of more exotic
non-Abelian anyons.

While our idealized setup does not take into account some ``real life''
complications, such as a finite stretch of the edge state (rather than its
single point) coupled to a vortex -- which will somewhat degrade the fidelity
of the information transfer -- our main goal has been to provide a ``proof of
principle". Further optimization of the transfer protocols with an eye on those
experimentally relevant effects should be a subject of future research.

\section*{Acknowledgments}
The authors would like to thank D.~Clarke and N.~Lindner for helpful
discussions. CYH and KS were supported in part by the DARPA-QuEST program. KS
was supported in part by NSF award DMR-1411359. CYH and GR  acknowledge the
support from the Packard foundation. The authors are also grateful to the IQIM,
an NSF center supported in part by the Moore foundation. In addition GR would
like to acknowledge the hospitality of the Aspen Center for Physics where part
of the work was done.

\appendix

\section{Solution of the equation of motion}
\label{app:solution-EOM}

Let us recall the Heisenberg equation of motion (EOM) in Eq.~\eqref{eq:EOM} for
the case of single chiral Majorana edge state:
\begin{subequations}
\label{eq:EOM-app}
\begin{eqnarray}
\label{eq:EOM-eta_SE-app}
(\partial_t +v_m \partial_x ) \eta &=& \frac{\alpha_1}{\hbar} \gamma_1  \delta(x-x_1)
+ \frac{\alpha_2}{\hbar} \gamma_2 \delta(x-x_2),
\\
\label{eq:EOM-gamma1_SE-app}
\partial_t \gamma_1 (t) &=& - \frac{\alpha_1(t)}{\hbar} \eta(x_1,t),
\\
\label{eq:EOM-gamma2_SE-app}
\partial_t \gamma_2 (t) &=& - \frac{\alpha_2(t)}{\hbar} \eta(x_2,t).
\end{eqnarray}
\end{subequations}
Here operators in Eqs.~\eqref{eq:EOM-app} are in the Heisenberg picture. The
goal is to solve these EOM, i.e., how these operators evolute as the function
of time given the initial operators, $\bar{\gamma}_a=\gamma_a(t_0)$ and
$\bar{\eta}(x,t_0)=\eta(x,t_0)$ at time $t= t_0 \to - \infty$.

To solve these EOMs in Eqs.~\eqref{eq:EOM}, we first observe that the retarded
Green functions for $\eta(x,t)$ and $\gamma_a(t)$ are given by
\begin{subequations}
\begin{eqnarray}
\label{eq:Green-function-eta}
G_{\eta}(x,t;x',t') &=& \Theta(t-t') \delta(x-x'- v_m (t-t')),
\\
\label{eq:Green-function-gamma}
G_{\gamma_a}(t,t') &=& \Theta(t-t'),
\end{eqnarray}
\end{subequations}
where $\Theta(t)$ is a Heaviside Theta function. We can then relate the
$\eta(x,t)$ field with Majorana modes $\gamma_a(t)$ from
Eq.~\eqref{eq:EOM-eta_SE-app} and give $\eta(x_a,t)$ as
\begin{eqnarray}
\label{eq:eta-x1-gamma}
\eta(x_1,t)&=&  \Theta\left(t-t_0 \right) \frac{\alpha_1 (t)}{2 \hbar v_m} \gamma_1(t),
\\
\label{eq:eta-x2-gamma}
\eta(x_2,t)&=& \Theta\left(t- t_0- \frac{\Delta x}{v_m} \right)
\frac{\alpha_1\! \left(t-{\Delta x}/{v_m}\right) }{\hbar v_m}\,\gamma_1\! \left(t-\frac{\Delta x}{v_m} \right)  \nonumber
\\
&& \qquad + \frac{\alpha_2 (t)}{2 \hbar v_m}\, \Theta\left(t-t_0 \right) \gamma_2(t),
\end{eqnarray}
where $\Delta x \equiv x_2-x_1>0$ and we have used $\Theta(0)\equiv 1/2$. On
the other hand, $\gamma_a(t)$ can be related to $\eta(x_a,t)$ by
Eqs.~\eqref{eq:EOM-gamma1_SE-app} and~\eqref{eq:EOM-gamma2_SE-app} as
\begin{equation}
\label{eq:gamma_a-eta_a}
\gamma_{a}(t) = - \int_{t_0}^{t} dt' \frac{\alpha_{a}(t')}{\hbar} \eta(x_a,t') dt'
\end{equation}

We can now obtain a set of coupled differential equations only for Majorana
modes by substituting Eqs.~\eqref{eq:eta-x1-gamma} and~\eqref{eq:eta-x2-gamma}
into Eqs.~\eqref{eq:EOM-gamma1_SE-app} and~\eqref{eq:EOM-gamma1_SE-app}
yielding
\begin{subequations}
\label{eq:diff-eq-gamma_a}
\begin{eqnarray}
\partial_t \gamma_1(t) &=& -  \Theta\left(t-t_0 \right)  \frac{\alpha_{1} (t)^2}{2 \hbar^2 v_m} \gamma_1(t),
\\
\partial_t \gamma_2(t) &=& -  \Theta\left(t-\left(t_0+\frac{\Delta x}{v_m}\right) \right) \frac{\alpha_2(t)
\alpha\!\left(t- \frac{\Delta x}{v_m}\right)}{\hbar^2 v_m} \gamma_1\!\left(t- \frac{\Delta x}{v_m}\right)
\nonumber
\\
&& -  \Theta\left(t-t_0 \right) \frac{\alpha_{2} (t)^2}{2 \hbar^2 v_m} \gamma_2(t).
\end{eqnarray}
\end{subequations}
In terms of initial operators $\gamma_{a}(t_0) \equiv \bar{\gamma}_{a}$, solutions of these differential equations give the first part of solution in Eq.~\eqref{eq:solution-gamma_a}, $K_a(t_0,t)$ and $W(t_0,t,\Delta x)$ in Eqs.~\eqref{eq:K_a} and~\eqref{eq:W_t}.

On the other hand, a decoupled integral-differential equation for $\eta(x,t)$
can be obtained similarly and is given by
\begin{multline}
\label{eq:diff-eq-eta}
(\partial_t + v_m \partial_x ) \eta(x,t)
\\
= - \frac{\alpha_{1}(t)}{\hbar^2} \Theta(t-t_0) \delta (x-x_1) \int_{t_0}^{t} dt' \alpha_1(t') \eta(x_1,t')
\\
- \frac{\alpha_{2}(t)}{\hbar^2} \Theta(t-t_0) \delta (x-x_2) \int_{t_0}^{t} dt' \alpha_2(t') \eta(x_2,t').
\end{multline}
As $\eta(x,t)$ is a chiral field, we can include the effect of couplings at $x=x_1$ and $x=x_2$ in sequence for solving this integral-differential equation. Therefore, let us consider the first part of the integral-differential equation
\begin{multline}
\label{eq:diff-eq-eta-app-1}
(\partial_t + v_m \partial_x ) \eta^{(1)}(x,t) =
\\
 - \frac{\alpha_{1}(t)}{\hbar^2} \Theta(t-t_0) \delta (x-x_1) \int_{t_0}^{t} dt' \alpha_1(t') \eta^{(1)}(x_1,t'),
\end{multline}
which implies the following formal solution
\begin{multline}
\label{eq:eta1-formal}
\eta^{(1)}(x,t)= \eta^{(0)}(x,t)
\\
- \Theta(x-x_1) \frac{\alpha_1(t-\frac{x-x_1}{v_m})}{\hbar^2 v_m} \int_{t_0}^{t-\frac{x-x_1}{v_m}} d \tau \alpha_1(\tau) \eta^{(1)}(x,\tau).
\end{multline}
Here, $\eta^{(0)}(x,t)\equiv \eta(x-v_m(t-t_0),t_0)$ origins from the chiral flow of the initial Majorana edge field.
Then, the full solution can be solved iteratively and can be formally expressed by
\begin{widetext}
\begin{multline}
\label{eq:eta1-formal-1}
\eta^{(1)}(x,t)= \eta^{(0)}(x,t) - \Theta(x-x_1) \frac{\alpha_1(t-\frac{x-x_1}{v_m})}{\hbar v_m} \times
\left\{ \int_{t_0}^{t-\frac{x-x_1}{v_m}} d \tau_0 \frac{\alpha_1(\tau_0)}{\hbar} \eta^{(0)}(x_1,\tau_0)  - \int_{t_0}^{t-\frac{x-x_1}{v_m}} d \tau_0 \frac{\alpha_1(\tau_0)^2}{\hbar^2 v_m} \int_{t_0}^{\tau_0} d \tau_1 \frac{\alpha_1(\tau_1)}{\hbar} \eta^{(0)}(x_1, \tau_1) \right.
\\
\left.
+ \int_{t_0}^{t-\frac{x-x_1}{v_m}} d \tau_0 \frac{\alpha_1(\tau_0)^2}{\hbar^2 v_m} \int_{t_0}^{\tau_0} d \tau_1 \frac{\alpha_1(\tau_1)^2}{\hbar^2 v_m} \int_{t_0}^{\tau_1} \frac{\alpha_{1}(\tau_2)}{\hbar} \eta^{(0)}(x_1,\tau_2) +\dots
\right\} .
\end{multline}
Now the order of integrations in each term can be rearranged with the following identity
\begin{equation}
\label{eq:exchange-integration}
\int_{t_0}^{A} d\tau_0 g(\tau_0) \int_{t_0}^{\tau_0} d \tau_1 f(\tau_1) = \int_{t_0}^{A} d \tau_1 f(\tau_1) \int_{\tau_1}^{A} d \tau_0 g(\tau_0)  \overset{\tau_0 \leftrightarrow \tau_1}{\xrightarrow{\hspace*{0.9cm}}}  \int_{t_0}^{A} d \tau_0 f(\tau_0) \int_{\tau_0}^{A} d \tau_1 g(\tau_1).
\end{equation}
\end{widetext}

After summing infinite terms, one obtains
\begin{multline}
\label{eq:eta1-formal-2}
\eta^{(1)}(x,t) =  \eta^{(0)}(x,t) - \Theta(x-x_1) \frac{\alpha_1\!
\left(t-{(x-x_1)}/{v_m}\right)}{\hbar^2 v_m}\\
\times \int_{t_0}^{t-\frac{x-x_1}{v_m}} d \tau_0\, \alpha_1(\tau_0)\,
K_1\!\left(\tau_0, t-\frac{x-x_1}{v_m}\right)\, \eta^{(0)}(x_1,\tau_0).
\end{multline}

We note that this solution of $\eta^{(1)}(x,t)$ can be fully related by initial
operators, $\eta(x,t_0)$.

To include the coupling at $x_2$ in Eq.~\eqref{eq:diff-eq-eta},  we can use
$\eta^{(1)}(x,t)$ as the incoming condition for Eq.~\eqref{eq:diff-eq-eta} at
$x=x_2$ because of the chirality of the EOM. Therefore, the full solution of
\eqref{eq:diff-eq-eta} is given by

\begin{multline}
\label{eq:eta-formal}
\eta(x,t) = \eta^{(1)}(x,t)
 - \Theta(x-x_2) \frac{\alpha_2\!\left(t-\frac{x-x_2}{v_m}\right)}{\hbar^2 v_m}\\
 \times \int_{t_0}^{t-\frac{x-x_2}{v_m}} d \tau_0\, \alpha_2(\tau_0)\,
 K_2\!\left(\tau_0,t-\frac{x-x_2}{v_m}\right)\, \eta^{(1)}(x_2,\tau_0) .
\end{multline}

Again, the solution of  $\eta(x,t)$ is expressed in terms of $\eta(x,t_0)$.
Finally, the contribution of initial operators $\eta(x,t_0)$ to $\gamma_a(t)$,
i.e. the second part (the integration part) of solutions in
Eqs.~\eqref{eq:solution-gamma_a}, can be obtained by employing
Eq.~\eqref{eq:gamma_a-eta_a} with solution in Eq.~\eqref{eq:eta-formal}.

\section{Edge-induced Polarization}
\label{app:polarization}

The goal of this Appendix is to describe the phenomenon of edge-induced
polarization, namely the non-zero expectation value $\langle
i\gamma_1(t)\gamma_2 (t)\rangle$ for two Majorana modes coupled to a chiral
Majorana single edge~\cite{Clarke2011b}. Here we will focus on the situation
where the coupling constants $\alpha_1$ and $\alpha_2$ are time independent and
will consider the limit $\Delta x =x_2 -x_1\to 0$ and $t\to \infty$. In this
limit, we observe that the contributions of initial operators $\bar{\gamma}_a$
to Eqs.~\eqref{eq:solution-gamma_a} decay exponentially and hence the only
important contributions are those resulting from the coupling to the chiral
edge state. Hence $\gamma_a$'s can be approximated as
\begin{equation}
\label{eq:solution-gamma_a-longtime}
\begin{split}
\gamma_1(t) \sim & - \alpha_1 \int_{t_0}^{t} d t'  K_{1} (t',t) \eta^{(0)} (x_1,t'),
\\
\gamma_2(t)\sim &  -\alpha_2 \int_{t_0}^{t} d t'  K_2(t',t) \eta^{(1)} (x_2,t').
\end{split}
\end{equation}
The qubit polarization is then formally expressed as
\begin{widetext}
\begin{equation}
\label{eq:polarization-formal}
\langle i\gamma_1(t)\gamma_2 (t)\rangle = i \frac{\alpha_1 \alpha_2}{\hbar^2} \left\langle   \int_{t_0}^{t} d t'  K_{1} (t',t) \eta^{(0)} (x_1,t') \int_{t_0}^{t} d t'' K_2(t'',t) \left( \eta^{(0)}(x_2,t'')
-\frac{\alpha_1^{2}}{\hbar^2 v_m}  \int_{t_0}^{t''} d \tau_0  K_1(\tau_0,t'') \eta^{(0)}(x_1,\tau_0) \right) \right\rangle,
\end{equation}
\end{widetext}
where we have used Eq.~\eqref{eq:eta1-formal-2} with $\Delta x\to0^{+}$ for
$\eta^{(1)} (x_2,t')$. We observe that the polarization are induced through two mechanisms: (I) polarization due to direct entanglement of Majorana modes and edge state, i.e., the first part of Eq.~\eqref{eq:polarization-formal}, and (II) polarization due to transfer quantum information of Majorana mode $\gamma_1$ to mode $\gamma_2$ through the edge, i.e. the second part of Eq.~\eqref{eq:polarization-formal}.

Before we evaluate the magnitude of the induced polarization, we first observe
that
\begin{equation}
\label{eq:Ka-constant}
K_a(t_i,t_f) =\exp \left[-\frac{\alpha_a^2}{2 \hbar^2 v_m}(t_f-t_i) \right]
\end{equation}
because $\alpha_a$ is time independent. We will also need the following correlation function
\begin{equation}
\label{eq:eta-eta-corr}
\langle \eta^{(0)}(x,t) \eta^{(0)} (y,t') \rangle= \frac{1}{i \pi}\, \frac{1}{(x-y) - v_m (t'-t'')} .
\end{equation}

Let us first focus on the contribution from mechanism (I),
\begin{multline}
\label{eq:poar-I}
\langle i\gamma_1(t)\gamma_2 (t)\rangle_{(I)} =
i \frac{\alpha_1 \alpha_2}{\hbar^2}  \int_{t_0}^{t} d t'  \int_{t_0}^{t} d t''   K_{1} (t',t)  K_2(t'',t)
\\
\times \left\langle \eta^{(0)} (x_1,t')   \eta^{(0)}(x_2,t'') \right\rangle.
\end{multline}
Using Eqs.~\eqref{eq:Ka-constant} and~\eqref{eq:eta-eta-corr} together with changes of variables $\tau'=t-t'$ and $\tau''=t-t''$ , we have
\begin{equation}
\label{eq:poar-I-inter-1}
- \frac{\alpha_1 \alpha_2}{\pi \hbar^2 v_m} \int^{\infty}_{0} d \tau'  \int^{\infty}_{0} d \tau'' \frac{\exp \left[ - \frac{\alpha_1^2}{2 \hbar^2 v_m} \tau' -\frac{\alpha_2^2}{2 \hbar^2 v_m} \tau'' \right]}{\tau''-\tau'}.
\end{equation}
Here we have taken $t-t_0\to \infty$. With a further change of variable $T=(\tau'+\tau'')/2$ and $\Delta \tau= (\tau''-\tau')/2$, the double integral can be carried out and gives
\begin{equation}
\label{eq:poar-I-final}
\langle i\gamma_1\gamma_2 (t\to \infty)\rangle_{(I)} = \frac{2 \alpha_1 \alpha_2 }{\pi (\alpha_2^2 +\alpha_1^2)} \ln\left(\frac{\alpha_2}{\alpha_1}\right)^2
\end{equation}

Now we turn to evaluate the contribution from mechanism (II),
\begin{multline}
\label{eq:poar-II}
\langle i\gamma_1(t)\gamma_2 (t)\rangle_{(II)} = - i \frac{\alpha_1^3 \alpha_2}{\hbar^4 v_m}    \int_{t_0}^{t} d t'  K_{1} (t',t) \int_{t_0}^{t} d t'' K_2(t'',t)
\\
\times \int_{t_0}^{t''} d \tau_0  K_1(\tau_0,t'')
\left\langle \eta^{(0)} (x_1,t')  \eta^{(0)}(x_1,\tau_0) \right\rangle.
\end{multline}
Exchanging the order of integral over $t''$ and $\tau_0$ and using
Eq.~\eqref{eq:exchange-integration} and explicitly integrate out $t''$, we
obtain
\begin{multline}
\label{eq:poar-II-inter-1}
 \frac{2\alpha_1^3 \alpha_2}{\pi \hbar^2 v_m(\alpha_2^2 -\alpha_1^2)} \int_{t_0}^{t} d t'  e^{ -\frac{\alpha_1^2}{2 \hbar^2 v_m}(t-t') }
 \\
 \times \int_{t_0}^{t} d \tau_0 \frac{ e^{\frac{\alpha_1^2}{2 \hbar^2 v_m}(\tau_0-t)} -e^{\frac{\alpha_2^2}{2 \hbar^2 v_m}(\tau_0-t)} }{t'-\tau_0}
\end{multline}
By performing changes of variables $\tau' =t-t'$ and $\tau''=t-\tau_0$ and taking $t-t_0\to \infty$, we then have
\begin{multline}
\label{eq:poar-II-inter-2}
 \frac{2\alpha_1^3 \alpha_2}{\pi \hbar^2 v_m(\alpha_2^2 -\alpha_1^2)} \int_{0}^{\infty} d \tau' d \tau'' e^{ -\frac{\alpha_1^2 \tau' }{2  \hbar^2 v_m}}  \frac{ e^{-\frac{\alpha_1^2 \tau''}{2 \hbar^2 v_m}} -e^{-\frac{\alpha_2^2 \tau''}{2 \hbar^2 v_m}} }{\tau''-\tau}.
\end{multline}
This integral is similar to that in Eq.~\eqref{eq:poar-I-inter-1}. Hence, with similar procedure, we can show that
\begin{equation}
\langle i\gamma_1\gamma_2 (t\to \infty)\rangle_{(II)} =   \frac{4\alpha_1^3 \alpha_2}{\pi (\alpha_2^2 -\alpha_1^2)(\alpha_2^2 -\alpha_1^2)} \ln \left(\frac{\alpha_2}{\alpha_1}\right)^2.
\end{equation}

By adding two part of contributions, the total polarization when two Majorana modes are coupled to the edge is give by
\begin{equation}
\langle i\gamma_1\gamma_2 (t\to \infty)\rangle =  \frac{2\alpha_1 \alpha_2}{\pi (\alpha_2^2 -\alpha_1^2)} \ln \left(\frac{\alpha_2}{\alpha_1}\right)^2.
\end{equation}
This result is consistent with what found in Ref.~\onlinecite{Clarke2011b} and
has its peak value $2/\pi$ for $\alpha_1/\alpha_2=1$.

\begin{widetext}
\section{Unequal velocities}
\label{app:unequal-velocity}

If the two edge states propagate with different velocities, the approach taken
in the main text, Sec.~\ref{sec:setup} and \ref{sec:QI-transfer}, is no longer
valid. One can not reduce the problem to that of a single edge state. Instead,
we can investigate directly the Green's function that enter the calculation of
$W_\text{2E}$ in Eq.~\eqref{eq:solution-gamma_a-2edge}. Starting with the
Hamiltonian of the two edges, we have:
\begin{equation}
H_\text{e} = \int dx \left\{ - i \frac{\hbar}{4} \left[v_L\eta^L (x) \partial_x \eta^L(x) +  v_R \eta^R (x) \partial_x \eta^R(x) \right]+ i m(x) \eta^{L}(x) \eta^{R}(x) \right\}.
\end{equation}
This can be rewritten in terms of Pauli matrices:
\begin{equation}
H_\text{e} = \int dx \left\{\left(\begin{array}{c} \eta^L \\ \eta^R \end{array}\right) ^{\dagger}\left[ - i \frac{\hbar}{4} \left( \overline{v}{\bf 1}+\delta v \sigma^z\right)\partial_x + m(x)\sigma^y\right]  \left(\begin{array}{c} \eta^L \\ \eta^R \end{array}\right)  \right\}.
\end{equation}
where we define $\delta v=\left({v_L-v_R}\right)/{2}$ and
$\overline{v}=\left({v_L+v_R}\right)/{2}$. The equations of motion for the
Green function are
\begin{equation}
\left[ \partial_t  + \left( \overline{v}{\bf 1}+\delta v \sigma^z\right)\partial_x - i \frac{m(x)}{\hbar}\sigma^y\right]  \left(\begin{array}{c} G_{\eta^L}(x-x',t-t') \\ G_{\eta^R} (x-x',t-t') \end{array}\right) ={\bf 1} \delta(x-x')\delta(t-t').
\end{equation}
Here, $x'$ and $t'$ are the source space and time. When $m(x)$ has no spatial
dependence, the Green's function is formally resolved in Fourier space as
\begin{equation}
G(x-x',t-t')= - i \int\frac{dk}{2\pi}\int\frac{d\omega}{2\pi} \frac{(\overline{v} k-\omega)-\delta v k \sigma^z + (m/\hbar) \sigma^y}{(\overline{v} k-\omega)^2-(\delta v k)^2-(m/\hbar)^2} e^{ik (x-x') -i \omega (t-t')}.
\end{equation}

By integrating over the frequency domain and choosing the pole that gives the
retarded Green's function, we obtain
\begin{multline}
G(x-x',t-t')= \Theta(t-t') \int\frac{dk}{2\pi} \frac{e^{i k (x-x')}}{2} \left[ \left(e^{-i \omega_+ (t-t')}+e^{-i \omega_- (t-t')}\right)
 +\frac{\hbar \delta v k\sigma^z}{\sqrt{m^2+\hbar^2k^2\delta v^2}} \left(e^{-i \omega_+ (t-t')} - e^{-i \omega_- (t-t')}\right) \right.
\\
\left.+
 \frac{m\sigma^y}{\sqrt{m^2+\hbar^2k^2\delta v^2}} \left(e^{-i \omega_- (t-t')}-e^{-i \omega_+ (t-t')}\right)
\right] ,
\label{keq}
\end{multline}
with $\omega_{\pm}=\overline{v} k\pm\sqrt{(m/\hbar)^2+ k^2\delta v^2}$. In the limit $\delta v \to 0$, the $2\times 2$ Green's function becomes
\begin{equation}
G(x-x',t-t')= \Theta(t-t') \delta((x-x')- \overline{v}(t-t')) \left( \cos \left(\frac{m}{\hbar} (t-t') \right){\bf 1} + i  \sin\left(\frac{m}{\hbar} (t-t')\right)  \sigma_y \right),
\label{keq-dv0}
\end{equation}
after taking the $k$ integration. As an alternative way, this Green's function
will lead to results obtained in Sec.~\ref{subsec:two-edge}.
\end{widetext}

Because the term that mixes the two edges plays important role for constructing
$W_\text{2E}$ in Eq. (\ref{eq:solution-gamma_a-2edge}), we shall focus on the
off-diagonal $\sigma_y$ term, $G_\text{y}(x-x',t-t')$. The role this function
plays becomes clear when we consider how to modify the result for the
all-important $W$ function from Eq. (\ref{eq:W_t}). The way our answer is
modified when the velocities are different is:
\begin{multline}
W(t_0,t,\Delta x) = - \int_{t_0}^{t} dt_1\int_{t_1}^{t}dt_2  \frac{\alpha_2(t_2) \alpha_1(t_1)}{\hbar^2 v_m}
\\ \times K_2(t',t_2) K_{1} (t_0,t_1)G_\text{y}(\Delta x,t_2-t_1),
\end{multline}
and $G_\text{y}(\Delta x,t_2-t_1)$ reduces to a $\sin\Delta\theta\cdot
\delta(t_2-t_1-\Delta x/v)$ in the equal-velocity case, with $\Delta \theta$
defined in Eq. (\ref{eq:Delta-theta}).

Let us next approximate $G_\text{y}(\Delta x,t_2-t_1)$. To allow information
transferring between the two edges, we must have $\frac{m \Delta x}{
\hbar\overline{v} } \sim 1$ and hence $\frac{ \hbar k \overline{v} }{m} \sim
1$. In the limit $\delta v \ll \overline{v}$, we can confine ourselves to the
case of $\hbar k\delta v\ll m$. Thus we can expand the expression in Eq.
(\ref{keq}) to the second order in $\delta v$ and obtain a Gaussian integral
\begin{multline}
G_\text{y}(x-x',t-t')\approx  \Theta(t-t') \int\frac{dk}{2\pi}\frac{\sigma^y}{2}e^{i k [(x-x')-\overline{v} (t-t')]}
\\
\times\left(e^{-\frac{1}{2}\left(\frac{\hbar \delta v
k}{m}\right)^2+i\left(\frac{m}{\hbar}+\frac{\hbar (k \delta
v)^2}{2m}\right)(t-t')}-e^{-\frac{1}{2}\left(\frac{\hbar \delta v
k}{m}\right)^2-i \left(\frac{m}{\hbar}+\frac{\hbar (k \delta
v)^2}{2m}\right)(t-t') }\right)
\\
=i \sigma^y  \Theta(t-t') \frac{m }{\sqrt{2 \pi} \hbar \delta v}\text{Im}\left[
\frac{e^{-\frac{1}{2}\left(\frac{m}{\hbar\delta v}\right)^2\frac{((x-x')-\overline{v}(t-t'))^2}
{1-i m (t-t')/\hbar}+i m (t-t')/\hbar }}{\sqrt{1- i\, m (t-t') / \hbar} }\right].
\label{keq-sdv}
\end{multline}

$G_\text{y}(\Delta x,t_2-t_1)$ has a time width of about $\delta t\sim
\frac{\hbar\delta v}{\overline{v}m}$ around $(x-x')-\overline{v}(t-t')=0$.
Hence the higher $\frac{m x}{\hbar\overline{v}}$ is, the more oscillatory the
function will be as a function of time at the second Majorana site. Observing
that Eq.~\eqref{keq-sdv} reduces to the result in Eq.~\eqref{keq-dv0} for
$\frac{\hbar\delta v}{m} \to 0$. As a result, we expect that the effect of the
velocity difference becomes negligible when $\frac{\hbar\delta v}{m} \ll \Delta
x$. In the limit  of our consideration, this condition is in general satisfied.
Indeed, carrying out the integral in Eq. (\ref{keq-sdv}) approximately yields a
simple suppression term of $e^{\left(-\delta v/v\right)^2}$ relative to the
results in Fig. \ref{fig:weight}.

\bibliographystyle{apsrev4-1}
\bibliography{kirref,espionage}

\begin{thebibliography}{36}%
\makeatletter
\providecommand \@ifxundefined [1]{%
 \@ifx{#1\undefined}
}%
\providecommand \@ifnum [1]{%
 \ifnum #1\expandafter \@firstoftwo
 \else \expandafter \@secondoftwo
 \fi
}%
\providecommand \@ifx [1]{%
 \ifx #1\expandafter \@firstoftwo
 \else \expandafter \@secondoftwo
 \fi
}%
\providecommand \natexlab [1]{#1}%
\providecommand \enquote  [1]{``#1''}%
\providecommand \bibnamefont  [1]{#1}%
\providecommand \bibfnamefont [1]{#1}%
\providecommand \citenamefont [1]{#1}%
\providecommand \href@noop [0]{\@secondoftwo}%
\providecommand \href [0]{\begingroup \@sanitize@url \@href}%
\providecommand \@href[1]{\@@startlink{#1}\@@href}%
\providecommand \@@href[1]{\endgroup#1\@@endlink}%
\providecommand \@sanitize@url [0]{\catcode `\\12\catcode `\$12\catcode
  `\&12\catcode `\#12\catcode `\^12\catcode `\_12\catcode `\%12\relax}%
\providecommand \@@startlink[1]{}%
\providecommand \@@endlink[0]{}%
\providecommand \url  [0]{\begingroup\@sanitize@url \@url }%
\providecommand \@url [1]{\endgroup\@href {#1}{\urlprefix }}%
\providecommand \urlprefix  [0]{URL }%
\providecommand \Eprint [0]{\href }%
\providecommand \doibase [0]{http://dx.doi.org/}%
\providecommand \selectlanguage [0]{\@gobble}%
\providecommand \bibinfo  [0]{\@secondoftwo}%
\providecommand \bibfield  [0]{\@secondoftwo}%
\providecommand \translation [1]{[#1]}%
\providecommand \BibitemOpen [0]{}%
\providecommand \bibitemStop [0]{}%
\providecommand \bibitemNoStop [0]{.\EOS\space}%
\providecommand \EOS [0]{\spacefactor3000\relax}%
\providecommand \BibitemShut  [1]{\csname bibitem#1\endcsname}%
\let\auto@bib@innerbib\@empty
%</preamble>
\bibitem [{\citenamefont {Leinaas}\ and\ \citenamefont
  {Myrheim}(1977)}]{Leinaas1977}%
  \BibitemOpen
  \bibfield  {author} {\bibinfo {author} {\bibfnamefont {J.~M.}\ \bibnamefont
  {Leinaas}}\ and\ \bibinfo {author} {\bibfnamefont {J.}~\bibnamefont
  {Myrheim}},\ }\href@noop {} {\bibfield  {journal} {\bibinfo  {journal} {Nuovo
  Cimento B}\ }\textbf {\bibinfo {volume} {37B}},\ \bibinfo {pages} {1}
  (\bibinfo {year} {1977})}\BibitemShut {NoStop}%
\bibitem [{\citenamefont {Wilczek}(1982{\natexlab{a}})}]{Wilczek1982a}%
  \BibitemOpen
  \bibfield  {author} {\bibinfo {author} {\bibfnamefont {F.}~\bibnamefont
  {Wilczek}},\ }\href {\doibase 10.1103/PhysRevLett.48.1144} {\bibfield
  {journal} {\bibinfo  {journal} {Phys. Rev. Lett.}\ }\textbf {\bibinfo
  {volume} {48}},\ \bibinfo {pages} {1144} (\bibinfo {year}
  {1982}{\natexlab{a}})}\BibitemShut {NoStop}%
\bibitem [{\citenamefont {Wilczek}(1982{\natexlab{b}})}]{Wilczek1982b}%
  \BibitemOpen
  \bibfield  {author} {\bibinfo {author} {\bibfnamefont {F.}~\bibnamefont
  {Wilczek}},\ }\href {\doibase 10.1103/PhysRevLett.49.957} {\bibfield
  {journal} {\bibinfo  {journal} {Phys. Rev. Lett.}\ }\textbf {\bibinfo
  {volume} {49}},\ \bibinfo {pages} {957} (\bibinfo {year}
  {1982}{\natexlab{b}})}\BibitemShut {NoStop}%
\bibitem [{\citenamefont {Arovas}\ \emph {et~al.}(1984)\citenamefont {Arovas},
  \citenamefont {Schrieffer},\ and\ \citenamefont {Wilczek}}]{Arovas1984}%
  \BibitemOpen
  \bibfield  {author} {\bibinfo {author} {\bibfnamefont {D.}~\bibnamefont
  {Arovas}}, \bibinfo {author} {\bibfnamefont {J.~R.}\ \bibnamefont
  {Schrieffer}}, \ and\ \bibinfo {author} {\bibfnamefont {F.}~\bibnamefont
  {Wilczek}},\ }\href {\doibase 10.1103/PhysRevLett.53.722} {\bibfield
  {journal} {\bibinfo  {journal} {Phys. Rev. Lett.}\ }\textbf {\bibinfo
  {volume} {53}},\ \bibinfo {pages} {722} (\bibinfo {year} {1984})}\BibitemShut
  {NoStop}%
\bibitem [{\citenamefont {Moore}\ and\ \citenamefont {Read}(1991)}]{Moore1991}%
  \BibitemOpen
  \bibfield  {author} {\bibinfo {author} {\bibfnamefont {G.}~\bibnamefont
  {Moore}}\ and\ \bibinfo {author} {\bibfnamefont {N.}~\bibnamefont {Read}},\
  }\href {\doibase doi:10.1016/0550-3213(91)90407-O} {\bibfield  {journal}
  {\bibinfo  {journal} {Nucl. Phys. B}\ }\textbf {\bibinfo {volume} {360}},\
  \bibinfo {pages} {362} (\bibinfo {year} {1991})}\BibitemShut {NoStop}%
\bibitem [{\citenamefont {Nayak}\ and\ \citenamefont
  {Wilczek}(1996)}]{Nayak1996c}%
  \BibitemOpen
  \bibfield  {author} {\bibinfo {author} {\bibfnamefont {C.}~\bibnamefont
  {Nayak}}\ and\ \bibinfo {author} {\bibfnamefont {F.}~\bibnamefont
  {Wilczek}},\ }\href@noop {} {\bibfield  {journal} {\bibinfo  {journal} {Nucl.
  Phys. B}\ }\textbf {\bibinfo {volume} {479}},\ \bibinfo {pages} {529}
  (\bibinfo {year} {1996})},\ \Eprint {http://arxiv.org/abs/cond-mat/9605145}
  {cond-mat/9605145} \BibitemShut {NoStop}%
\bibitem [{\citenamefont {Nayak}\ \emph {et~al.}(2008)\citenamefont {Nayak},
  \citenamefont {Simon}, \citenamefont {Stern}, \citenamefont {Freedman},\ and\
  \citenamefont {Das~Sarma}}]{Nayak2008}%
  \BibitemOpen
  \bibfield  {author} {\bibinfo {author} {\bibfnamefont {C.}~\bibnamefont
  {Nayak}}, \bibinfo {author} {\bibfnamefont {S.~H.}\ \bibnamefont {Simon}},
  \bibinfo {author} {\bibfnamefont {A.}~\bibnamefont {Stern}}, \bibinfo
  {author} {\bibfnamefont {M.}~\bibnamefont {Freedman}}, \ and\ \bibinfo
  {author} {\bibfnamefont {S.}~\bibnamefont {Das~Sarma}},\ }\href {\doibase
  10.1103/RevModPhys.80.1083} {\bibfield  {journal} {\bibinfo  {journal} {Rev.
  Mod. Phys.}\ }\textbf {\bibinfo {volume} {80}},\ \bibinfo {eid} {1083}
  (\bibinfo {year} {2008})},\ \Eprint {http://arxiv.org/abs/arXiv:0707.1889}
  {arXiv:0707.1889} \BibitemShut {NoStop}%
\bibitem [{\citenamefont {Kitaev}(2003)}]{Kitaev2003}%
  \BibitemOpen
  \bibfield  {author} {\bibinfo {author} {\bibfnamefont {A.~Y.}\ \bibnamefont
  {Kitaev}},\ }\href {\doibase 10.1016/S0003-4916(02)00018-0} {\bibfield
  {journal} {\bibinfo  {journal} {Ann. Phys.}\ }\textbf {\bibinfo {volume}
  {303}},\ \bibinfo {pages} {2} (\bibinfo {year} {2003})},\ \Eprint
  {http://arxiv.org/abs/quant-ph/9707021} {quant-ph/9707021} \BibitemShut
  {NoStop}%
\bibitem [{\citenamefont {Clarke}\ and\ \citenamefont
  {Shtengel}(2011)}]{Clarke2011b}%
  \BibitemOpen
  \bibfield  {author} {\bibinfo {author} {\bibfnamefont {D.~J.}\ \bibnamefont
  {Clarke}}\ and\ \bibinfo {author} {\bibfnamefont {K.}~\bibnamefont
  {Shtengel}},\ }\href {\doibase 10.1088/1367-2630/13/5/055005} {\bibfield
  {journal} {\bibinfo  {journal} {New J. Phys.}\ }\textbf {\bibinfo {volume}
  {13}},\ \bibinfo {pages} {055005} (\bibinfo {year} {2011})},\ \Eprint
  {http://arxiv.org/abs/arXiv:1102.2016} {arXiv:1102.2016} \BibitemShut
  {NoStop}%
\bibitem [{\citenamefont {Hassler}\ \emph {et~al.}(2010)\citenamefont
  {Hassler}, \citenamefont {Akhmerov}, \citenamefont {Hou},\ and\ \citenamefont
  {Beenakker}}]{Hassler2010a}%
  \BibitemOpen
  \bibfield  {author} {\bibinfo {author} {\bibfnamefont {F.}~\bibnamefont
  {Hassler}}, \bibinfo {author} {\bibfnamefont {A.~R.}\ \bibnamefont
  {Akhmerov}}, \bibinfo {author} {\bibfnamefont {C.-Y.}\ \bibnamefont {Hou}}, \
  and\ \bibinfo {author} {\bibfnamefont {C.~W.~J.}\ \bibnamefont {Beenakker}},\
  }\href {\doibase 10.1088/1367-2630/12/12/125002} {\bibfield  {journal}
  {\bibinfo  {journal} {New J. Phys.}\ }\textbf {\bibinfo {volume} {12}},\
  \bibinfo {pages} {125002} (\bibinfo {year} {2010})},\ \Eprint
  {http://arxiv.org/abs/arXiv:1005.3423} {arXiv:1005.3423} \BibitemShut
  {NoStop}%
\bibitem [{\citenamefont {Bonderson}\ and\ \citenamefont
  {Lutchyn}(2011)}]{Bonderson2011b}%
  \BibitemOpen
  \bibfield  {author} {\bibinfo {author} {\bibfnamefont {P.}~\bibnamefont
  {Bonderson}}\ and\ \bibinfo {author} {\bibfnamefont {R.~M.}\ \bibnamefont
  {Lutchyn}},\ }\href {\doibase 10.1103/PhysRevLett.106.130505} {\bibfield
  {journal} {\bibinfo  {journal} {Phys. Rev. Lett.}\ }\textbf {\bibinfo
  {volume} {106}},\ \bibinfo {pages} {130505} (\bibinfo {year} {2011})},\
  \Eprint {http://arxiv.org/abs/arXiv:1011.1784} {arXiv:1011.1784} \BibitemShut
  {NoStop}%
\bibitem [{\citenamefont {Hassler}\ \emph {et~al.}(2011)\citenamefont
  {Hassler}, \citenamefont {Akhmerov},\ and\ \citenamefont
  {Beenakker}}]{Hassler2011}%
  \BibitemOpen
  \bibfield  {author} {\bibinfo {author} {\bibfnamefont {F.}~\bibnamefont
  {Hassler}}, \bibinfo {author} {\bibfnamefont {A.~R.}\ \bibnamefont
  {Akhmerov}}, \ and\ \bibinfo {author} {\bibfnamefont {C.~W.~J.}\ \bibnamefont
  {Beenakker}},\ }\href {\doibase 10.1088/1367-2630/13/9/095004} {\bibfield
  {journal} {\bibinfo  {journal} {New J. Phys.}\ }\textbf {\bibinfo {volume}
  {13}},\ \bibinfo {pages} {095004} (\bibinfo {year} {2011})},\ \Eprint
  {http://arxiv.org/abs/arXiv:1105.0315} {arXiv:1105.0315} \BibitemShut
  {NoStop}%
\bibitem [{\citenamefont {Pekker}\ \emph {et~al.}(2013)\citenamefont {Pekker},
  \citenamefont {Hou}, \citenamefont {Manucharyan},\ and\ \citenamefont
  {Demler}}]{Pekker2013a}%
  \BibitemOpen
  \bibfield  {author} {\bibinfo {author} {\bibfnamefont {D.}~\bibnamefont
  {Pekker}}, \bibinfo {author} {\bibfnamefont {C.-Y.}\ \bibnamefont {Hou}},
  \bibinfo {author} {\bibfnamefont {V.~E.}\ \bibnamefont {Manucharyan}}, \ and\
  \bibinfo {author} {\bibfnamefont {E.}~\bibnamefont {Demler}},\ }\href
  {\doibase 10.1103/PhysRevLett.111.107007} {\bibfield  {journal} {\bibinfo
  {journal} {Phys. Rev. Lett.}\ }\textbf {\bibinfo {volume} {111}},\ \bibinfo
  {pages} {107007} (\bibinfo {year} {2013})},\ \Eprint
  {http://arxiv.org/abs/arXiv:1301.3161} {arXiv:1301.3161} \BibitemShut
  {NoStop}%
\bibitem [{\citenamefont {Kovalev}\ \emph {et~al.}(2013)\citenamefont
  {Kovalev}, \citenamefont {De},\ and\ \citenamefont
  {Shtengel}}]{Kovalev2014a}%
  \BibitemOpen
  \bibfield  {author} {\bibinfo {author} {\bibfnamefont {A.~A.}\ \bibnamefont
  {Kovalev}}, \bibinfo {author} {\bibfnamefont {A.}~\bibnamefont {De}}, \ and\
  \bibinfo {author} {\bibfnamefont {K.}~\bibnamefont {Shtengel}},\ }\href
  {\doibase 10.1103/PhysRevLett.112.106402} {\bibfield  {journal} {\bibinfo
  {journal} {Phys. Rev. Lett.}\ }\textbf {\bibinfo {volume} {112}},\ \bibinfo
  {pages} {106402} (\bibinfo {year} {2013})},\ \Eprint
  {http://arxiv.org/abs/arXiv:1306.2339} {arXiv:1306.2339} \BibitemShut
  {NoStop}%
\bibitem [{\citenamefont {Volovik}(1999)}]{Volovik1999}%
  \BibitemOpen
  \bibfield  {author} {\bibinfo {author} {\bibfnamefont {G.}~\bibnamefont
  {Volovik}},\ }\href {\doibase 10.1134/1.568223} {\bibfield  {journal}
  {\bibinfo  {journal} {JETP Letters}\ }\textbf {\bibinfo {volume} {70}},\
  \bibinfo {pages} {609} (\bibinfo {year} {1999})},\ \bibinfo {note}
  {[\textit{Pis'ma Zh. Eksp. Teor. Fiz.} \textbf{70}, 601 (1999)]}\BibitemShut
  {NoStop}%
\bibitem [{\citenamefont {Read}\ and\ \citenamefont {Green}(2000)}]{Read2000}%
  \BibitemOpen
  \bibfield  {author} {\bibinfo {author} {\bibfnamefont {N.}~\bibnamefont
  {Read}}\ and\ \bibinfo {author} {\bibfnamefont {D.}~\bibnamefont {Green}},\
  }\href {\doibase 10.1103/PhysRevB.61.10267} {\bibfield  {journal} {\bibinfo
  {journal} {Phys. Rev. B}\ }\textbf {\bibinfo {volume} {61}},\ \bibinfo
  {pages} {10267} (\bibinfo {year} {2000})},\ \Eprint
  {http://arxiv.org/abs/cond-mat/9906453} {cond-mat/9906453} \BibitemShut
  {NoStop}%
\bibitem [{\citenamefont {Ivanov}(2001)}]{Ivanov2001}%
  \BibitemOpen
  \bibfield  {author} {\bibinfo {author} {\bibfnamefont {D.~A.}\ \bibnamefont
  {Ivanov}},\ }\href {\doibase 10.1103/PhysRevLett.86.268} {\bibfield
  {journal} {\bibinfo  {journal} {Phys. Rev. Lett.}\ }\textbf {\bibinfo
  {volume} {86}},\ \bibinfo {pages} {268} (\bibinfo {year} {2001})},\ \Eprint
  {http://arxiv.org/abs/cond-mat/0005069} {cond-mat/0005069} \BibitemShut
  {NoStop}%
\bibitem [{\citenamefont {Alicea}(2012)}]{Alicea2012a}%
  \BibitemOpen
  \bibfield  {author} {\bibinfo {author} {\bibfnamefont {J.}~\bibnamefont
  {Alicea}},\ }\href {\doibase 10.1088/0034-4885/75/7/076501} {\bibfield
  {journal} {\bibinfo  {journal} {Rep. Prog. Phys.}\ }\textbf {\bibinfo
  {volume} {75}},\ \bibinfo {pages} {076501} (\bibinfo {year} {2012})},\
  \Eprint {http://arxiv.org/abs/arXiv:1202.1293} {arXiv:1202.1293} \BibitemShut
  {NoStop}%
\bibitem [{\citenamefont {Beenakker}(2013)}]{Beenakker2013a}%
  \BibitemOpen
  \bibfield  {author} {\bibinfo {author} {\bibfnamefont {C.~W.~J.}\
  \bibnamefont {Beenakker}},\ }\href {\doibase
  10.1146/annurev-conmatphys-030212-184337} {\bibfield  {journal} {\bibinfo
  {journal} {Annu. Rev. Condens. Matter Phys.}\ }\textbf {\bibinfo {volume}
  {4}},\ \bibinfo {pages} {113} (\bibinfo {year} {2013})},\ \Eprint
  {http://arxiv.org/abs/arXiv:1112.1950} {arXiv:1112.1950} \BibitemShut
  {NoStop}%
\bibitem [{\citenamefont {Mourik}\ \emph {et~al.}(2012)\citenamefont {Mourik},
  \citenamefont {Zuo}, \citenamefont {Frolov}, \citenamefont {Plissard},
  \citenamefont {Bakkers},\ and\ \citenamefont {Kouwenhoven}}]{Mourik2012}%
  \BibitemOpen
  \bibfield  {author} {\bibinfo {author} {\bibfnamefont {V.}~\bibnamefont
  {Mourik}}, \bibinfo {author} {\bibfnamefont {K.}~\bibnamefont {Zuo}},
  \bibinfo {author} {\bibfnamefont {S.~M.}\ \bibnamefont {Frolov}}, \bibinfo
  {author} {\bibfnamefont {S.~R.}\ \bibnamefont {Plissard}}, \bibinfo {author}
  {\bibfnamefont {E.~P. A.~M.}\ \bibnamefont {Bakkers}}, \ and\ \bibinfo
  {author} {\bibfnamefont {L.~P.}\ \bibnamefont {Kouwenhoven}},\ }\href
  {\doibase 10.1126/science.1222360} {\bibfield  {journal} {\bibinfo  {journal}
  {Science}\ }\textbf {\bibinfo {volume} {336}},\ \bibinfo {pages} {1003}
  (\bibinfo {year} {2012})}\BibitemShut {NoStop}%
\bibitem [{\citenamefont {Deng}\ \emph {et~al.}(2012)\citenamefont {Deng},
  \citenamefont {Yu}, \citenamefont {Huang}, \citenamefont {Larsson},
  \citenamefont {Caroff},\ and\ \citenamefont {Xu}}]{Deng2012}%
  \BibitemOpen
  \bibfield  {author} {\bibinfo {author} {\bibfnamefont {M.~T.}\ \bibnamefont
  {Deng}}, \bibinfo {author} {\bibfnamefont {C.~L.}\ \bibnamefont {Yu}},
  \bibinfo {author} {\bibfnamefont {G.~Y.}\ \bibnamefont {Huang}}, \bibinfo
  {author} {\bibfnamefont {M.}~\bibnamefont {Larsson}}, \bibinfo {author}
  {\bibfnamefont {P.}~\bibnamefont {Caroff}}, \ and\ \bibinfo {author}
  {\bibfnamefont {H.~Q.}\ \bibnamefont {Xu}},\ }\href {\doibase
  10.1021/nl303758w} {\bibfield  {journal} {\bibinfo  {journal} {Nano Lett.}\
  }\textbf {\bibinfo {volume} {12}},\ \bibinfo {pages} {6414} (\bibinfo {year}
  {2012})},\ \Eprint {http://arxiv.org/abs/arXiv:1204.4130} {arXiv:1204.4130}
  \BibitemShut {NoStop}%
\bibitem [{\citenamefont {Das}\ \emph {et~al.}(2012)\citenamefont {Das},
  \citenamefont {Ronen}, \citenamefont {Most}, \citenamefont {Oreg},
  \citenamefont {Heiblum},\ and\ \citenamefont {Shtrikman}}]{Das2012a}%
  \BibitemOpen
  \bibfield  {author} {\bibinfo {author} {\bibfnamefont {A.}~\bibnamefont
  {Das}}, \bibinfo {author} {\bibfnamefont {Y.}~\bibnamefont {Ronen}}, \bibinfo
  {author} {\bibfnamefont {Y.}~\bibnamefont {Most}}, \bibinfo {author}
  {\bibfnamefont {Y.}~\bibnamefont {Oreg}}, \bibinfo {author} {\bibfnamefont
  {M.}~\bibnamefont {Heiblum}}, \ and\ \bibinfo {author} {\bibfnamefont
  {H.}~\bibnamefont {Shtrikman}},\ }\href {\doibase 10.1038/nphys2479}
  {\bibfield  {journal} {\bibinfo  {journal} {Nat. Phys.}\ }\textbf {\bibinfo
  {volume} {8}},\ \bibinfo {pages} {887} (\bibinfo {year} {2012})},\ \Eprint
  {http://arxiv.org/abs/arXiv:1205.7073} {arXiv:1205.7073} \BibitemShut
  {NoStop}%
\bibitem [{\citenamefont {Rokhinson}\ \emph {et~al.}(2012)\citenamefont
  {Rokhinson}, \citenamefont {Liu},\ and\ \citenamefont
  {Furdyna}}]{Rokhinson2012}%
  \BibitemOpen
  \bibfield  {author} {\bibinfo {author} {\bibfnamefont {L.~P.}\ \bibnamefont
  {Rokhinson}}, \bibinfo {author} {\bibfnamefont {X.}~\bibnamefont {Liu}}, \
  and\ \bibinfo {author} {\bibfnamefont {J.~K.}\ \bibnamefont {Furdyna}},\
  }\href {\doibase 10.1038/nphys2429} {\bibfield  {journal} {\bibinfo
  {journal} {Nat. Phys.}\ }\textbf {\bibinfo {volume} {8}},\ \bibinfo {pages}
  {795} (\bibinfo {year} {2012})},\ \Eprint
  {http://arxiv.org/abs/arXiv:1204.4212} {arXiv:1204.4212} \BibitemShut
  {NoStop}%
\bibitem [{\citenamefont {Nadj-Perge}\ \emph {et~al.}(2014)\citenamefont
  {Nadj-Perge}, \citenamefont {Drozdov}, \citenamefont {Li}, \citenamefont
  {Chen}, \citenamefont {Jeon}, \citenamefont {Seo}, \citenamefont {MacDonald},
  \citenamefont {Bernevig},\ and\ \citenamefont {Yazdani}}]{Nadj-Perge2014}%
  \BibitemOpen
  \bibfield  {author} {\bibinfo {author} {\bibfnamefont {S.}~\bibnamefont
  {Nadj-Perge}}, \bibinfo {author} {\bibfnamefont {I.~K.}\ \bibnamefont
  {Drozdov}}, \bibinfo {author} {\bibfnamefont {J.}~\bibnamefont {Li}},
  \bibinfo {author} {\bibfnamefont {H.}~\bibnamefont {Chen}}, \bibinfo {author}
  {\bibfnamefont {S.}~\bibnamefont {Jeon}}, \bibinfo {author} {\bibfnamefont
  {J.}~\bibnamefont {Seo}}, \bibinfo {author} {\bibfnamefont {A.~H.}\
  \bibnamefont {MacDonald}}, \bibinfo {author} {\bibfnamefont {B.~A.}\
  \bibnamefont {Bernevig}}, \ and\ \bibinfo {author} {\bibfnamefont
  {A.}~\bibnamefont {Yazdani}},\ }\href {\doibase 10.1126/science.1259327}
  {\bibfield  {journal} {\bibinfo  {journal} {Science}\ }\textbf {\bibinfo
  {volume} {346}},\ \bibinfo {pages} {602} (\bibinfo {year} {2014})},\ \Eprint
  {http://arxiv.org/abs/arXiv:1410.0682} {arXiv:1410.0682} \BibitemShut
  {NoStop}%
\bibitem [{\citenamefont {Kitaev}(2001)}]{Kitaev2001}%
  \BibitemOpen
  \bibfield  {author} {\bibinfo {author} {\bibfnamefont {A.~Y.}\ \bibnamefont
  {Kitaev}},\ }\href {\doibase 10.1070/1063-7869/44/10S/S29} {\bibfield
  {journal} {\bibinfo  {journal} {Phys.-Usp.}\ }\textbf {\bibinfo {volume}
  {44}},\ \bibinfo {pages} {131} (\bibinfo {year} {2001})},\ \Eprint
  {http://arxiv.org/abs/cond-mat/0010440} {cond-mat/0010440} \BibitemShut
  {NoStop}%
\bibitem [{\citenamefont {Lutchyn}\ \emph {et~al.}(2010)\citenamefont
  {Lutchyn}, \citenamefont {Sau},\ and\ \citenamefont
  {Das~Sarma}}]{Lutchyn2010a}%
  \BibitemOpen
  \bibfield  {author} {\bibinfo {author} {\bibfnamefont {R.~M.}\ \bibnamefont
  {Lutchyn}}, \bibinfo {author} {\bibfnamefont {J.~D.}\ \bibnamefont {Sau}}, \
  and\ \bibinfo {author} {\bibfnamefont {S.}~\bibnamefont {Das~Sarma}},\ }\href
  {\doibase 10.1103/PhysRevLett.105.077001} {\bibfield  {journal} {\bibinfo
  {journal} {Phys. Rev. Lett.}\ }\textbf {\bibinfo {volume} {105}},\ \bibinfo
  {pages} {077001} (\bibinfo {year} {2010})},\ \Eprint
  {http://arxiv.org/abs/arXiv:1002.4033} {arXiv:1002.4033} \BibitemShut
  {NoStop}%
\bibitem [{\citenamefont {Oreg}\ \emph {et~al.}(2010)\citenamefont {Oreg},
  \citenamefont {Refael},\ and\ \citenamefont {von Oppen}}]{Oreg2010}%
  \BibitemOpen
  \bibfield  {author} {\bibinfo {author} {\bibfnamefont {Y.}~\bibnamefont
  {Oreg}}, \bibinfo {author} {\bibfnamefont {G.}~\bibnamefont {Refael}}, \ and\
  \bibinfo {author} {\bibfnamefont {F.}~\bibnamefont {von Oppen}},\ }\href
  {\doibase 10.1103/PhysRevLett.105.177002} {\bibfield  {journal} {\bibinfo
  {journal} {Phys. Rev. Lett.}\ }\textbf {\bibinfo {volume} {105}},\ \bibinfo
  {pages} {177002} (\bibinfo {year} {2010})},\ \Eprint
  {http://arxiv.org/abs/arXiv:1003.1145} {arXiv:1003.1145} \BibitemShut
  {NoStop}%
\bibitem [{\citenamefont {Nadj-Perge}\ \emph {et~al.}(2013)\citenamefont
  {Nadj-Perge}, \citenamefont {Drozdov}, \citenamefont {Bernevig},\ and\
  \citenamefont {Yazdani}}]{Nadj-Perge2013}%
  \BibitemOpen
  \bibfield  {author} {\bibinfo {author} {\bibfnamefont {S.}~\bibnamefont
  {Nadj-Perge}}, \bibinfo {author} {\bibfnamefont {I.~K.}\ \bibnamefont
  {Drozdov}}, \bibinfo {author} {\bibfnamefont {B.~A.}\ \bibnamefont
  {Bernevig}}, \ and\ \bibinfo {author} {\bibfnamefont {A.}~\bibnamefont
  {Yazdani}},\ }\href {\doibase 10.1103/PhysRevB.88.020407} {\bibfield
  {journal} {\bibinfo  {journal} {Phys. Rev. B}\ }\textbf {\bibinfo {volume}
  {88}},\ \bibinfo {pages} {020407} (\bibinfo {year} {2013})},\ \Eprint
  {http://arxiv.org/abs/arXiv:1303.6363} {arXiv:1303.6363} \BibitemShut
  {NoStop}%
\bibitem [{\citenamefont {Bravyi}(2006)}]{Bravyi2006}%
  \BibitemOpen
  \bibfield  {author} {\bibinfo {author} {\bibfnamefont {S.}~\bibnamefont
  {Bravyi}},\ }\href {\doibase 10.1103/PhysRevA.73.042313} {\bibfield
  {journal} {\bibinfo  {journal} {Phys. Rev. A}\ }\textbf {\bibinfo {volume}
  {73}},\ \bibinfo {pages} {042313} (\bibinfo {year} {2006})},\ \Eprint
  {http://arxiv.org/abs/quant-ph/0511178} {quant-ph/0511178} \BibitemShut
  {NoStop}%
\bibitem [{\citenamefont {Kwon}\ \emph {et~al.}(2004)\citenamefont {Kwon},
  \citenamefont {Sengupta},\ and\ \citenamefont {Yakovenko}}]{Kwon2004}%
  \BibitemOpen
  \bibfield  {author} {\bibinfo {author} {\bibfnamefont {H.-J.}\ \bibnamefont
  {Kwon}}, \bibinfo {author} {\bibfnamefont {K.}~\bibnamefont {Sengupta}}, \
  and\ \bibinfo {author} {\bibfnamefont {V.}~\bibnamefont {Yakovenko}},\ }\href
  {\doibase 10.1140/epjb/e2004-00066-4} {\bibfield  {journal} {\bibinfo
  {journal} {Eur. Phys. J. B}\ }\textbf {\bibinfo {volume} {37}},\ \bibinfo
  {pages} {349} (\bibinfo {year} {2004})},\ \Eprint
  {http://arxiv.org/abs/cond-mat/0210148} {cond-mat/0210148} \BibitemShut
  {NoStop}%
\bibitem [{\citenamefont {Serban}\ \emph {et~al.}(2010)\citenamefont {Serban},
  \citenamefont {B\'eri}, \citenamefont {Akhmerov},\ and\ \citenamefont
  {Beenakker}}]{Serban2010}%
  \BibitemOpen
  \bibfield  {author} {\bibinfo {author} {\bibfnamefont {I.}~\bibnamefont
  {Serban}}, \bibinfo {author} {\bibfnamefont {B.}~\bibnamefont {B\'eri}},
  \bibinfo {author} {\bibfnamefont {A.~R.}\ \bibnamefont {Akhmerov}}, \ and\
  \bibinfo {author} {\bibfnamefont {C.~W.~J.}\ \bibnamefont {Beenakker}},\
  }\href {\doibase 10.1103/PhysRevLett.104.147001} {\bibfield  {journal}
  {\bibinfo  {journal} {Phys. Rev. Lett.}\ }\textbf {\bibinfo {volume} {104}},\
  \bibinfo {pages} {147001} (\bibinfo {year} {2010})}\BibitemShut {NoStop}%
\bibitem [{\citenamefont {Yao}\ \emph {et~al.}(2011)\citenamefont {Yao},
  \citenamefont {Laumann}, \citenamefont {Gorshkov}, \citenamefont {Weimer},
  \citenamefont {Jiang}, \citenamefont {Cirac}, \citenamefont {Zoller},\ and\
  \citenamefont {Lukin}}]{Yao2011}%
  \BibitemOpen
  \bibfield  {author} {\bibinfo {author} {\bibfnamefont {N.~Y.}\ \bibnamefont
  {Yao}}, \bibinfo {author} {\bibfnamefont {C.~R.}\ \bibnamefont {Laumann}},
  \bibinfo {author} {\bibfnamefont {A.~V.}\ \bibnamefont {Gorshkov}}, \bibinfo
  {author} {\bibfnamefont {H.}~\bibnamefont {Weimer}}, \bibinfo {author}
  {\bibfnamefont {L.}~\bibnamefont {Jiang}}, \bibinfo {author} {\bibfnamefont
  {J.~I.}\ \bibnamefont {Cirac}}, \bibinfo {author} {\bibfnamefont
  {P.}~\bibnamefont {Zoller}}, \ and\ \bibinfo {author} {\bibfnamefont {M.~D.}\
  \bibnamefont {Lukin}},\ }\href {\doibase 10.1038/ncomms2531} {\bibfield
  {journal} {\bibinfo  {journal} {Nature Comm.}\ }\textbf {\bibinfo {volume}
  {4}},\ \bibinfo {pages} {1585} (\bibinfo {year} {2011})},\ \Eprint
  {http://arxiv.org/abs/arXiv:1110.3788} {arXiv:1110.3788} \BibitemShut
  {NoStop}%
\bibitem [{\citenamefont {Rosenow}\ \emph {et~al.}(2008)\citenamefont
  {Rosenow}, \citenamefont {Halperin}, \citenamefont {Simon},\ and\
  \citenamefont {Stern}}]{Rosenow2008a}%
  \BibitemOpen
  \bibfield  {author} {\bibinfo {author} {\bibfnamefont {B.}~\bibnamefont
  {Rosenow}}, \bibinfo {author} {\bibfnamefont {B.~I.}\ \bibnamefont
  {Halperin}}, \bibinfo {author} {\bibfnamefont {S.~H.}\ \bibnamefont {Simon}},
  \ and\ \bibinfo {author} {\bibfnamefont {A.}~\bibnamefont {Stern}},\ }\href
  {\doibase 10.1103/PhysRevLett.100.226803} {\bibfield  {journal} {\bibinfo
  {journal} {Phys. Rev. Lett.}\ }\textbf {\bibinfo {volume} {100}},\ \bibinfo
  {pages} {226803} (\bibinfo {year} {2008})},\ \Eprint
  {http://arxiv.org/abs/arXiv:0707.4474} {arXiv:0707.4474} \BibitemShut
  {NoStop}%
\bibitem [{\citenamefont {Rosenow}\ \emph {et~al.}(2009)\citenamefont
  {Rosenow}, \citenamefont {Halperin}, \citenamefont {Simon},\ and\
  \citenamefont {Stern}}]{Rosenow2009a}%
  \BibitemOpen
  \bibfield  {author} {\bibinfo {author} {\bibfnamefont {B.}~\bibnamefont
  {Rosenow}}, \bibinfo {author} {\bibfnamefont {B.~I.}\ \bibnamefont
  {Halperin}}, \bibinfo {author} {\bibfnamefont {S.~H.}\ \bibnamefont {Simon}},
  \ and\ \bibinfo {author} {\bibfnamefont {A.}~\bibnamefont {Stern}},\ }\href
  {\doibase 10.1103/PhysRevB.80.155305} {\bibfield  {journal} {\bibinfo
  {journal} {Phys. Rev. B}\ }\textbf {\bibinfo {volume} {80}},\ \bibinfo
  {pages} {155305} (\bibinfo {year} {2009})},\ \Eprint
  {http://arxiv.org/abs/arXiv:0906.0310} {arXiv:0906.0310} \BibitemShut
  {NoStop}%
\bibitem [{\citenamefont {Grosfeld}\ and\ \citenamefont
  {Stern}(2011)}]{Grosfeld2011}%
  \BibitemOpen
  \bibfield  {author} {\bibinfo {author} {\bibfnamefont {E.}~\bibnamefont
  {Grosfeld}}\ and\ \bibinfo {author} {\bibfnamefont {A.}~\bibnamefont
  {Stern}},\ }\href {\doibase 10.1073/pnas.1101469108} {\ \textbf {\bibinfo
  {volume} {108}},\ \bibinfo {pages} {11810} (\bibinfo {year}
  {2011})}\BibitemShut {NoStop}%
\bibitem [{\citenamefont {Wen}(1991)}]{Wen1991}%
  \BibitemOpen
  \bibfield  {author} {\bibinfo {author} {\bibfnamefont {X.~G.}\ \bibnamefont
  {Wen}},\ }\href@noop {} {\bibfield  {journal} {\bibinfo  {journal} {Phys.
  Rev. B}\ }\textbf {\bibinfo {volume} {43}},\ \bibinfo {pages} {11025}
  (\bibinfo {year} {1991})}\BibitemShut {NoStop}%
\end{thebibliography}%

\end{document}